\documentclass[12pt]{article}

\usepackage{latexsym}

\usepackage[font=small,labelfont=bf]{caption}

\usepackage{graphics}
\usepackage{graphicx}
\usepackage{epstopdf}
\usepackage{amssymb}
\usepackage{tabularx}
\usepackage{caption}
\usepackage{subcaption}
\usepackage{hyperref}
\usepackage{tabularx}

\graphicspath{{./Figures/}}
\begin{document}


\title {Quasi-periodic oscillations from the accretion disk around rotating traversable wormholes}

\author{
Efthimia Deligianni$^{1}$\footnote{E-mail: \texttt{efthimia.deligianni@uni-oldenburg.de}}, \, Jutta Kunz$^{1}$\footnote{E-mail: \texttt{jutta.kunz@uni-oldenburg.de}}, \, Petya Nedkova$^{1,2}$\footnote{E-mail: \texttt{pnedkova@phys.uni-sofia.bg}}, \\
Stoytcho  Yazadjiev$^{2,3}$ \footnote{E-mail: \texttt{yazad@phys.uni-sofia.bg}}, \, Radostina Zheleva$^{2}$\footnote{E-mail: \texttt{radostinazheleva94@gmail.com}}\\ \\
    {\footnotesize${}^{1}$  Institute of Physics, Carl von Ossietzky University of Oldenburg,}\\
    {\footnotesize 26111 Oldenburg, Germany}\\
    {\footnotesize${}^{2}$ Faculty of Physics, Sofia University,}\\
    {\footnotesize    5 James Bourchier Boulevard, Sofia~1164, Bulgaria }\\
    {\footnotesize${}^{3}$  Institute of Mathematics and Informatics,}\\
    {\footnotesize  Bulgarian Academy of Sciences, Acad. G. Bonchev 8,}\\
    {\footnotesize  Sofia 1113, Bulgaria}}
\date{}
\maketitle

\begin{abstract}
We study the quasi-periodic oscillations from the accretion disk around rotating traversable wormholes by means of the resonance models. We investigate the linear stability of the circular geodesic orbits in the equatorial plane for a general class of wormhole geometries deriving analytical expressions for the epicyclic frequencies. Since wormholes can often mimic black holes in the astrophysical observations, we analyze the properties of the quasi-circular oscillatory motion in comparison with the Kerr black hole. We demonstrate that wormholes possess distinctive features, which can be observationally significant. It is characteristic for the Kerr black hole that the orbital and the epicyclic frequencies obey a constant ordering in the whole range of the spin parameter. In contrast, for wormhole spacetimes we can have various types of orderings between the frequencies in the different regions of the parametric space. This enables the excitation of much more diverse types of resonances including parametric and forced resonances of lower order, which could lead to stronger observable signals. In addition, for co-rotating orbits the resonances can be excited in a very close neighbourhood of the wormhole throat for a wide range of values of the angular momentum, making wormholes a valuable laboratory for testing strong gravity.
\end{abstract}

\section{Introduction}

X-ray spectroscopy is a promising tool for testing gravity in the strong-field regime by studying the electromagnetic emission from the accretion disks around compact objects. Various features of the X-ray flux will be measured with a high precision by means of the next generation of X-ray satellites like LOFT \cite{Feroci:2016}, eXTP \cite{Zhang:2016}, or STROBE-X \cite{Wilson-Hodge:2017}, in particular  the quasi-periodic oscillations (QPOs) of the accretion flow. The quasi-periodic oscillations were experimentally detected in the X-ray flux from a number of low-mass binaries including neutron stars or black holes, as well as a few supermassive active galactic nuclei. They represent a number of characteristic peaks appearing in the X-ray spectrum from the compact object, including a low-frequency (Hz) signal and a couple of high-frequency (kHz) oscillations.

The precise physical mechanism for the formation of the quasi-periodic oscillations is currently unknown but some features suggest that they are a hydrodynamical phenomenon, rather than a manifestation of kinematic effects in the accretion disk like the Doppler modulation of fluxes from isolated hot spots. Such an example  is the discovery of the correlation between the low-and high-frequency QPOs, causing that their ratio remains stable among various X-ray sources \cite{Psaltis}-\cite{Mauche}. This motivated the development of  disk seismological models which explain the QPOs by means of certain trapped modes of the disk oscillations \cite{Kato:1980}-\cite{Kato:2004b}. It was further observed that the high-frequency QPOs scale inversely with the mass of the compact object and that the twin peaks follow a constant $3:2$ ratio. Thus, we got indications that the high-frequency QPOs are caused by relativistic effects so that they represent a suitable probe of the background spacetime. The characteristic frequency ratio of two integers further suggests that the source of the quasi-periodic oscillations can be some non-linear resonance process taking place in the inner disk.

Using the thin disk approximation resonance models were developed which can give explanation for the observed high-frequency QPOs \cite{Abramowicz:2001}-\cite{Torok:2005a}. Assuming that the fluid lines in the accretion disk follow nearly circular geodesic trajectories located in a single plane, we can associate two epicyclic frequencies with their dynamics. They describe the fluctuations from the perfectly circular motion in radial and vertical directions, respectively. In linear approximation the radial and vertical fluctuations can be considered independent and represented as two decoupled harmonic oscillations. However, non-linear effects cause interactions between the two oscillation modes and different types of resonances can be excited when the epicyclic frequencies or linear combinations of them scale as the ratio of two integer numbers. Depending on the physical process taking place in the accretion disk, resonances can be caused also by the coupling of the orbital and one of the epicyclic motions.

The quasi-periodic oscillations were studied as a tool for testing gravitational theories in a number of recent works \cite{Stella:1999}-\cite{Bambi:2012}. The properties of the epicyclic frequencies depend strongly on the underlying spacetime and can lead to observational effects, which can differentiate between the alternative theories of gravity in the strong field regime. In some cases the interaction of the compact object with the astrophysical surroundings should be also taken into account since it can lead to similar effects \cite{Nedkova:2020}. On the other hand, if the observational data is modeled correctly, the QPOs can provide evidence for the existence of more exotic compact objects like wormholes and naked singularities. Some investigations of broader classes of compact objects were performed including studies of the Kerr naked singularity \cite{Torok:2005b}, the Zipoy-Vorhees spacetime \cite{Malafarina:2019}, and the Tomimatsu-Sato solution \cite{Stefanov:2013}. The aim of our work is to study the quasi-periodic oscillations for a class of rotating traversable wormholes  by applying  the resonance models and to estimate some of their signatures, which can be important for the observations.

Wormholes are one of the significant predictions of the gravitational theories, which is still not confirmed observationally. In classical general relativity the construction of traversable wormholes is problematic since it requires the violation of the null energy condition. However, in quantum gravity this issue would be solved since quantum fields can provide in a natural way the necessary negative energy density. Alternatively, traversable wormholes arise in some modified theories of gravity  like in the Gauss-Bonnet theory or $f(R)$ theories, where the energy conditions are violated by the gravitational interaction itself without the need of additional matter fields \cite{Hochberg}-\cite{Kunz:2020b}. They can also exist in a mixed system with another compact object like a boson or neutron star \cite{Hoffmann}-\cite{Folomeev}. Thus, the idea that traversable wormholes can form in nature is reasonably well supported by theoretical arguments, and one of the goals of the next generation gravitational experiments is to search for evidence for their existence.

Various works investigated the possible observational signatures of wormholes. They include studies of the shadow \cite{Nedkova:2013}, \cite{Gyulchev:2018}, \cite{Bambi}, gravitational lensing \cite{Perlick}, \cite{Tsukamoto},  Lense-Thirring precession \cite{Chakraborty},  accretion disk radiation \cite{Harko}, iron line profile \cite{Zhou}, and quasinormal modes \cite{Kunz:2018}. It was demonstrated that in some phenomena wormholes can mimic closely the Kerr black hole, while in others they  possess distinctive features. Since it is hard to distinguish wormholes from black holes in certain experiments, being familiar with a broader set of characteristic effects will be useful for their identification.

The aim of our work is to provide further observable effects in the electromagnetic spectrum, which can distinguish wormholes from other compact objects, by studying the high-frequency quasi-periodic oscillations in wormhole spacetimes.  For the purpose we consider the class of geometries constructed by Teo which describes a general stationary and axisymmetric traversable wormhole \cite{Teo}. Teo's rotating wormhole geometry is a generalization of the Morris-Thorne static spherically symmetric wormhole following the same idea. Instead of searching for a solution of the gravitational field equations with a particular stress-energy tensor, the most general stationary and axisymmetric geometry is derived which describes a regular spacetime with wormhole topology. Any traversable wormhole solution within classical and semi-classical gravitational theories would belong to this class, so although there is a lot of generality in choosing the particular metric functions, the geometry is useful for modeling some wormhole properties which would be common irrespective of the gravitational theory. We should further note that it is hard to obtain exact solutions, which describe completely regular rotating wormholes. Even when coupling the Einstein equations to very simple matter fields like a phantom scalar field for example, only perturbative or numerical wormhole solutions are known \cite{Sushkov:2008a}-\cite{Kunz:2021}. Thus, while completely regular exact solutions are typically unavailable when rotation is included, the Teo's geometry provides the opportunity to investigate the properties of the rotating traversable wormholes in an analytical way, and gain some useful insights.

The paper is organized as follows. In section 2 we briefly describe the Teo's geometry considering in particular a class of metrics with integrable geodesic equations. In section 3 and 4 we study the existence and the linear stability of the circular geodesics in the equatorial plane and derive general expressions for their kinematic characteristics, and the radial and vertical epicyclic frequencies. Some properties which are valid for the whole class of traversable wormholes, which we consider, are also discussed. In section 5 we apply our results to a particular wormhole solution, which illustrates  certain characteristic effects for the wormhole geometries. In particular we investigate the behaviour of the epicyclic frequencies for different angular momenta,  and the possible types of ordering between the orbital and the epicyclic frequencies in the various regions of the parametric space. In section 6  we analyse the quasi-periodic oscillations  within the resonance models in comparison with the Kerr black hole. We demonstrate that in wormhole geometries resonances possess a richer structure, which leads to more diverse possibilities for modeling the observational data from the accretion disk spectroscopy. In the last section we give our conclusions.

\section{Rotating traversable wormholes}

The general class of geometries, which describes a stationary axisymmetric traversable wormhole was obtained by Teo in the form \cite{Teo}

\begin{equation}\label{metric}
    ds^2 = -N^2dt^2 + \left(1-\frac{b}{r}\right)^{-1}dr^2+r^2K^2\left[d\theta^2+ \sin^2\theta\left(d\phi-\omega dt\right)^2\right],
   \end{equation}
where all the metric functions depend only on the spherical coordinates $r$ and $\theta$, and are regular on the symmetry axis $\theta=0,\pi$. The metric function $N$ is connected to the gravitational redshift, $K$ is a measure of the radial distance with respect to the coordinate origin, $\omega$ is associated with the wormhole rotation, while $b$ determines the shape of  the wormhole throat.

In order to ensure the geometrical characteristics of the wormhole and its traversability the metric functions should satisfy certain requirements. We can construct a geometry, which contains a wormhole throat, if we consider the coordinate range $r\geq b$ and require that the metric possesses an apparent singularity at  $b=r$, where  the metric function $g_{rr}$ becomes divergent. In this way the wormhole throat is located at a constant radius $r=r_0$. If the  shape function does not depend on $\theta$ on the wormhole throat, i.e  $\partial_\theta b\left(r,\theta\right)=0$, the scalar curvature would be regular, so there would be no curvature singularity. In addition the redshift function should be finite and non-zero in all the coordinate range in order to avoid further curvature singularities or event horizons, while the proper radial distance should be a  positive and nondecreasing function. We can ensure the characteristic form of the wormhole by considering the embedding of its constant $t$ and $\theta$ cross-sections in a three dimensional Euclidian space. In this way the so called "flare-out" condition is derived, which reduces to the requirement that $db/dr<1$ at the wormhole throat.

The constructed geometry describes two identical regions, in each of which the radial coordinate takes the range $r\in [r_0, \infty)$, joined together at the wormhole throat $r=r_0$. We can further require that the regions are asymptotically flat. Then the metric functions should possess the following expansions at the spacetime infinity

 \begin{eqnarray}
     N &=& 1-\frac{M}{r}+\mathcal O\left(\frac{1}{r^2}\right), \quad~~~ K = 1+\mathcal O\left(\frac{1}{r}\right),\quad~~~\frac{b}{r} =\mathcal O\left(\frac{1}{r}\right),  \nonumber \\[2mm]
     \omega &=& \frac{2J}{r^3}+\mathcal O\left(\frac{1}{r^4}\right),
   \end{eqnarray}
where $M$ and $J$ are constants representing the ADM mass and the angular momentum of the wormhole, respectively.

There is a broad range of functions which can satisfy the described conditions and lead to a regular wormhole geometry. In our work we will consider the class of the geometries, where the metric functions $N$, $K$, $b$ and $\omega$  depend only on the radial coordinate. This case is of particular physical importance since the geodesic equations, which determine the particle and light propagation become integrable. In order to illustrate some characteristic effects for the wormhole geometries, we choose a particular metric which is simple enough, but still representative for the class. We set the shape function and radial distance function to be equal to constants, while for the rest of the metric functions we assume

\begin{eqnarray}\label{wormhole}
 N = \exp\left(-\frac{r_0}{r}\right), \quad~~~  \omega =\frac{2J}{r^3}, \quad~~~b=r_0, \quad~~~ K=1.
 \end{eqnarray}

Thus, the mass of the wormhole is equal to $M = r_0$. We can further introduce a spin parameter $a=J/M^2$ and represent the metric in a dimensionless form by making the conformal transformation and the rescaling

\begin{eqnarray}
dS^{2}= r_0^{-2}ds^{2}, \quad~~~t\rightarrow r_0t, \quad~~~r\rightarrow r_0r. \nonumber
\end{eqnarray}
In this way we obtain a wormhole solution with a unit mass and a throat located at $r=1$.

\section{Circular orbits in the equatorial plane}

For any stationary and axisymmetric metric we can derive some general expressions, which determine the kinematic quantities on the circular orbits in the equatorial plane. Let us consider the general form of the metric

\begin{eqnarray}
ds^2=g_{tt}\,dt^2+2g_{t\phi}\,dt d\phi+g_{rr}\,dr^2+g_{\theta\theta}\,d\theta^2+g_{\phi\phi}\,d\phi^2\,,
\end{eqnarray}
and denote the specific energy and angular momentum of the particles by $E$ and $L$, respectively. Since we have two Killing vectors with respect to time translations and rotations around the symmetry axis, the energy and the angular momentum are conserved on the geodesics. In addition, on each geodesic trajectory we have the constraint $g_{\mu\nu}{\dot x}^{\mu}{\dot x}^{\nu} =\epsilon$, where $\epsilon$ takes the value $\epsilon = -1$ for timelike geodesics, and $\epsilon = 0$ for null geodesics.  Choosing an affine parameter $\tau$ we can express the timelike geodesic equations in the equatorial plane as

\begin{eqnarray}
\frac{dt}{d\tau}&=&\frac{E g_{\phi\phi}+L g_{t\phi}}{g_{t\phi}^2-g_{tt}g_{\phi\phi}},   \nonumber   \\[2mm]
\frac{d\phi}{d\tau}&=&-\frac{E g_{t\phi}+ L g_{tt}}{g_{t\phi}^2-g_{tt}g_{\phi\phi}}, \nonumber \\[2mm]
    g_{rr}\left(\frac{dr}{d\tau}\right)^2&=&-1+\frac{E^2 g_{\phi\phi}+2 E L g_{t\phi}+ L^2g_{tt}}{g_{t\phi}^2-g_{tt}g_{\phi\phi}}.
    \label{geodeqs3}
\end{eqnarray}

In the last equation we can introduce an effective potential $V_{eff}$ given by

\begin{eqnarray}
V_{eff} = -1+\frac{E^2 g_{\phi\phi}+2 E L g_{t\phi} + L^2g_{tt}}{g_{t\phi}^2-g_{tt}g_{\phi\phi}}.
\end{eqnarray}

Then, the qualitative behavior of the radial motion is determined completely by the properties of the effective potential. In particular, circular orbits correspond to its stationary points

\begin{eqnarray}
V_{eff}(r)=0, \quad~~~ V_{eff,r}(r)=0,
\end{eqnarray}
where the comma denotes the derivative with respect to the radial coordinate. Solving this system of equations we can obtain the specific energy and angular momentum on the circular orbits in the form
\begin{eqnarray}
E &=&-\frac{g_{tt}+g_{t\phi}\omega_0}{\sqrt{-g_{tt}-2g_{t\phi}\omega_0-g_{\phi\phi}\omega_0^2}},    \label{Ecirc}  \\[2mm]
L &=& \frac{g_{t\phi}+g_{\phi\phi}\omega_0}{\sqrt{-g_{tt}-2g_{t\phi}\omega_0-g_{\phi\phi}\omega_0^2}},     \label{Lcirc}
\end{eqnarray}
in terms of the angular velocity

\begin{eqnarray}
\omega_0&=&\frac{d\phi}{dt}=\frac{-g_{t\phi,r}\pm \sqrt{(g_{t\phi,r})^2-g_{tt,r}g_{\phi\phi,r}}}{g_{\phi\phi,r}},
     \label{Omega}
\end{eqnarray}
where the $+/-$ sign refers to the co-rotating/counter-rotating orbits, respectively.

Timelike circular orbits exist in the region where the energy and the angular momentum are well-defined. The curves where they diverge correspond to the location of the null circular orbits, or the photon rings. They give the boundary of the region of existence of the timelike circular orbits. If only an unstable photon ring is present, the region of existence of the timelike circular orbits is simply connected, and for asymptotically flat spacetimes the location of the photon ring gives its lower limit in radial direction. If there are multiple photon rings, the region of existence can consist of several disjoined pieces, i.e. circular orbits will be located in several annular regions with gaps in between.

We calculate the kinematic quantities for the traversable wormhole spacetime ($\ref{metric}$)  obtaining the expressions 

\begin{eqnarray}\label{omega_0}
\omega_0 &=& \omega +\frac{r^2K^2\omega_{,r} \pm \sqrt{N^2_{,r}(r^2K^2)_{,r} + r^4K^4(\omega_{,r})^2}}{(r^2K^2)_{,r}},  \\[2mm] 
E &=& \frac{N^2 + r^2K^2\omega(\omega_0-\omega) }{\sqrt{N^2 - r^2K^2(\omega_0 -\omega)^2}}, \nonumber \\[2mm]
L &=& \frac{r^2K^2(\omega_0 -\omega)}{\sqrt{N^2 - r^2K^2(\omega_0 -\omega)^2}}, \nonumber
\end{eqnarray}
where we have the $+/-$ sign for co-rotating/counter-rotating orbits. Timelike circular orbits exist in the region where the inequality $N^2 - r^2K^2(\omega_0 -\omega)^2>0$ is satisfied.

In fig. $\ref{fig:stability}$ we demonstrate the domain of existence of the circular orbits for the particular wormhole solution given by ($\ref{wormhole}$). The rotating wormhole solution exists for any values of the spin parameter. However, since our aim is to make comparison with the Kerr black hole, we constrain its range to the limits $a\in[0,1]$. In order to illustrate the behavior of the counter-rotating circular orbits we include negative values of the spin parameter $a\in[-1,0)$. Thus, the co-rotating geodesics are represented by the region with positive values of $a$, while the region with the negative spin describes the counter-rotating ones. We see that the co-rotating circular orbits exist in the whole spacetime up to the wormhole throat. On the other hand, counter-rotating particles are expelled by the spinning compact object and can reach only to a certain radial distance, leaving a region around the wormhole throat where the counter-rotating circular orbits are not allowed.

\section{Stability of the circular orbits in the equatorial plane}

We consider the geodesic equations in the equatorial plane

\begin{equation}\label{geodesics}
\ddot{x^{\alpha}}+\Gamma^{\alpha}_{\,\,\,\beta\gamma}\dot{x^{\beta}}\dot{x^{\gamma}}=0\,,
\end{equation}
and perform a small perturbation from the circular motion $\tilde{x}^{\mu}(s) = x^{\mu}(s) + \xi^\mu(s)$, where $x^{\mu}(s)$  denotes the circular orbit and  $s$ is an affine parameter on the geodesic. Working in the linear approximation we can obtain the following system for the deviation $\xi^\mu(s)$ \cite{Aliev:1981}, \cite{Aliev:1986}
\begin{eqnarray}\label{pert}
&&\frac{d^2\xi^\mu}{dt^2} + 2\gamma^\mu_\alpha\frac{d\xi^\alpha}{dt} + \xi^b\partial_b{\cal V}^{\mu} = 0\, , \quad b = r, \theta \nonumber \\[2mm]
&&\gamma^\mu_\alpha =\left[\Gamma^\mu_{\alpha\beta} u^\beta(u^0)^{-1}\right]_{\theta=\pi/2}\, , \nonumber \\[2mm]
&& {\cal V}^{\mu} = \left[\gamma^\mu_\alpha u^\alpha(u^0)^{-1}\right]_{\theta=\pi/2},
\end{eqnarray}
where $\omega_0$ is the orbital frequency, and $u^\mu = \dot{x^\mu}= u^0(1, 0, 0, \omega_0)$ is the 4-velocity vector. For convenience we further introduce a separate notation for the cyclic coordinates $t$ and $\phi$ denoting them with capital Latin indices, while small Latin indices refer to the $r$ and $\theta$ coordinates. We can integrate directly the equations for the $t$ and $\phi$ perturbations obtaining
\begin{eqnarray}
\frac{d\xi^A}{dt} + 2\gamma^A_\alpha\xi^\alpha = 0\, , \quad A = t,\phi,
\end{eqnarray}
and substitute these expressions in the remaining part of the system. Considering the class of traversable wormholes with metric functions $N$, $K$, $b$ and $\omega$ depending only on the radial coordinate, we can show that the equations for the radial and vertical perturbations decouple and reduce to the form
\begin{eqnarray}\label{freq}
&&\frac{d^2\xi^r}{dt^2} + \omega_r^2\xi^r = 0\, , \label{pertx} \\[2mm]
&&\frac{d^2\xi^\theta}{dt^2} + \omega_\theta^2\xi^\theta = 0\, ,  \nonumber
\end{eqnarray}
where we have introduced the quantities
\begin{eqnarray}
&&\omega_r^2 = \partial_r {\cal V}^r - 4\gamma^r_A\gamma^A_r, \\[2mm] \nonumber
&&\omega_\theta^2 = \partial_\theta{\cal V}^\theta. \nonumber
\end{eqnarray}

This representation allows us to make conclusions about the linear stability of the circular motion in the equatorial plane. It is determined by the sign of the functions $\omega_r^2$ and $\omega_\theta^2$, which depend on the radial coordinate and the spin of the wormhole. When they are positive, the system ($\ref{freq}$) describes two harmonic oscillations around the circular  orbit in radial and vertical directions  with frequencies $\omega_r$ and $\omega_\theta$, which are called epicyclic frequencies. Then, the circular motion is stable in linear approximation. If one of  functions $\omega_r^2$ and $\omega_\theta^2$ becomes negative, a small perturbation from the circular orbit in the corresponding direction will deviate exponentially from it, and the circular motion gets unstable.

We apply the derived expressions to obtain the epicyclic frequencies for the rotating traversable wormholes 
\begin{eqnarray}\label{omega_xy}
\omega^2_\theta&=& (\omega_0 - \omega)^2\, ,  \\[2mm]
\omega^2_r&=& \frac{(b-r)}{rN^2}\left[\omega^2_\theta\, r^4 K^4 \omega_{,r}^2 + r K^2(\omega_0-\omega)(2r (N^2\omega_{,r})_{,r}-3N^2(r\omega_{,r})_{,r})\right]\nonumber \\
&& + \omega^2_\theta\,(b-r)\left[\frac{K}{r}(r^2 K_{,r})_{,r} - 3 K_{,r}(rK)_{,r}\right]  \nonumber \\
&&-\frac{(b-r)}{r^2}\left[3N N_{,r} + rN N_{,rr} - 3r(N_{,r})^2\right]\, ,  
\end{eqnarray}
where we restrict ourselves to the class of solutions with metric functions $N$, $K$, $b$ and $\omega$ depending only on $r$, and the orbital frequency $\omega_0$ is given by eq. $(\ref{omega_0})$.

From these expressions we can deduce some general properties of the considered wormhole geometries. We see that for all the wormhole solutions the vertical epicyclic frequency is always positive so the circular orbits are always stable with respect to vertical perturbations. Thus, the linear stability is determined only by the radial epicyclic frequency similar to the case of the Kerr black hole. In the static limit we obtain that the vertical epicyclic frequency coincides with the orbital frequency similar to the Schwarzschild black hole. In this way the circular motion is characterized by only two independent quantities.

\begin{figure}[t!]
            \setlength{\tabcolsep}{ 0 pt }{\small\tt
		           \begin{tabular}{ cc}
           \includegraphics[width=0.5\textwidth]{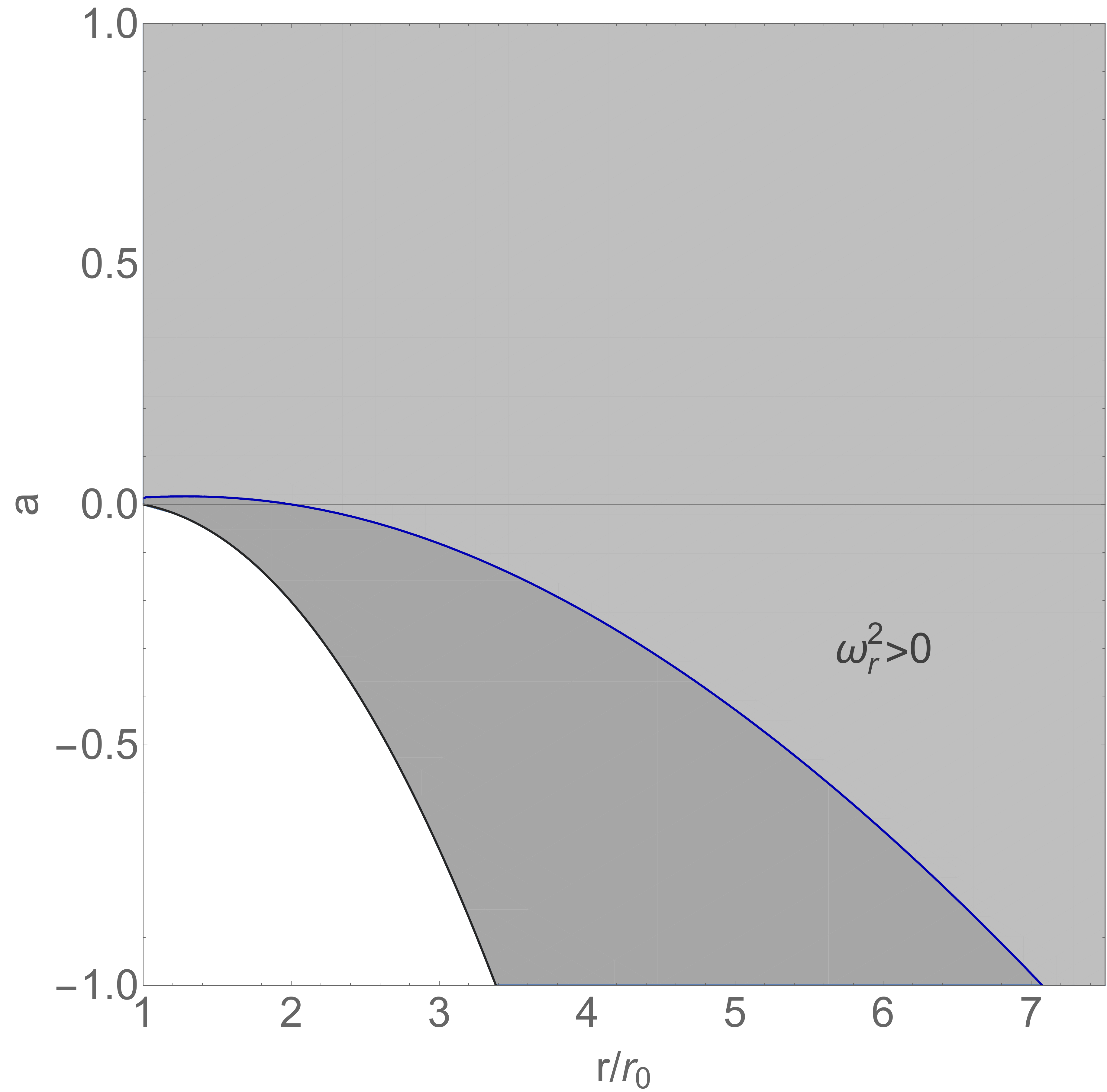}
		   \includegraphics[width=0.5\textwidth]{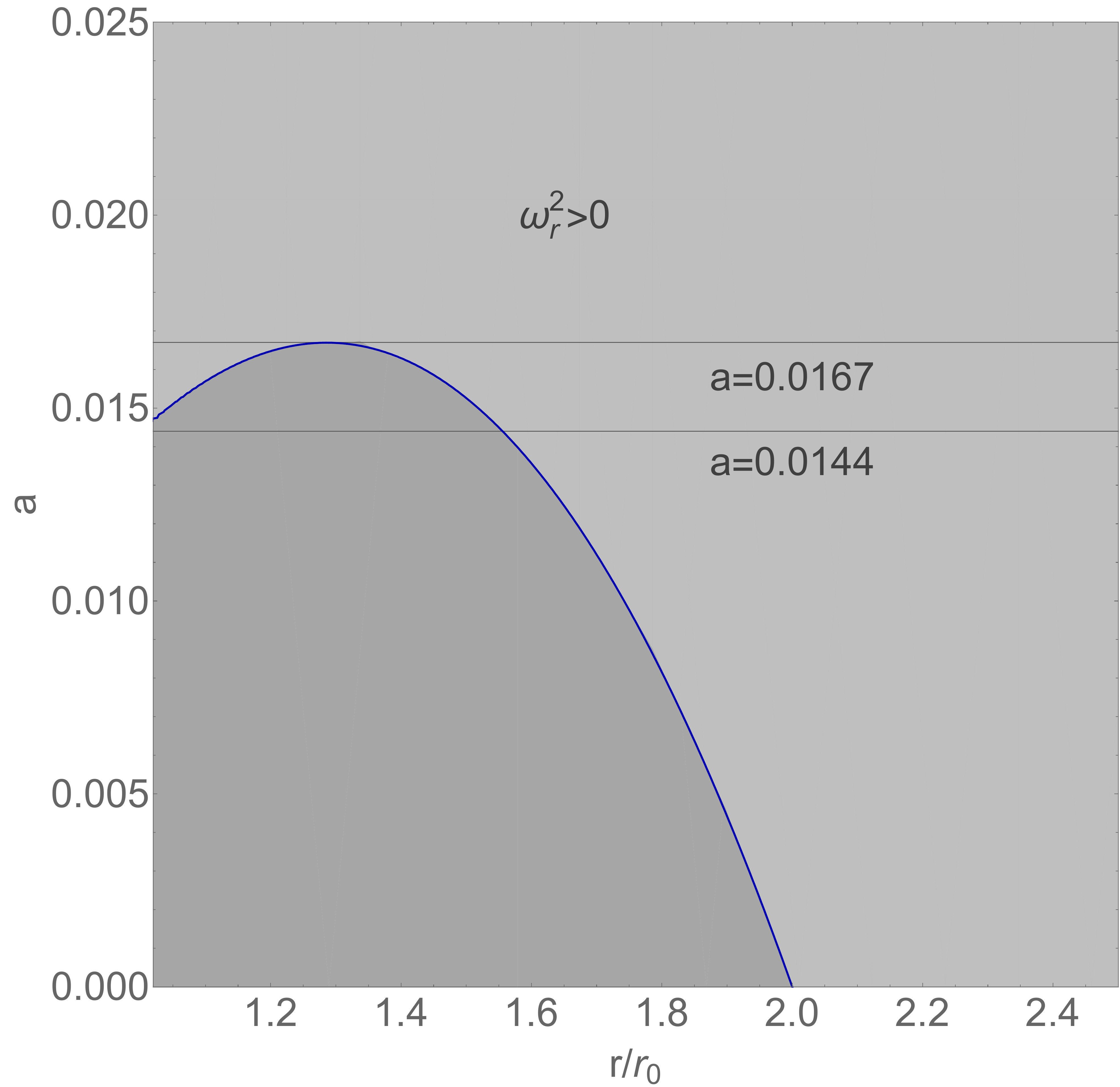} \\[1mm]
           \hspace{1.2cm}  $a)$ \hspace{7cm}  $b)$
        \end{tabular}}
		    \caption{\label{fig:stability}\small Existence and stability of the circular orbits in the equatorial plane for rotating traversable wormholes. The grey curve denotes the location of the null circular orbits, while on the blue curve the relation $\omega_r^2=0$ is satisfied. Timelike circular orbits exist in the grey region, and they are stable in the light grey one bounded by the blue curve. The positive values of the spin parameter represent the co-rotating circular orbits, while the negative ones correspond to the counter-rotating ones. }
\end{figure}

In order to get further intuition about the behavior of the circular orbits, we investigate the region of stability for the particular wormhole solution given by eq. ($\ref{wormhole}$). In fig. $\ref{fig:stability}$ we present the curve on which the radial epicyclic frequency vanishes, thus delimiting the region of stability of the circular orbits. In the region above the curve and on its righthand side the inequality $\omega_r^2>0$ is satisfied, so this part of the spacetime represents the region of stability of the timelike circular orbits in the equatorial plane.

We see that for most of the spin parameters in the range $a\in [0,1]$ the co-rotating orbits are stable in the whole spacetime. For small angular momenta of the wormhole  however, we get a qualitatively different situation. The curve $\omega_r^2=0$ possesses a maximum at $a=0.0167$ and intersects the wormhole throat at $a=0.0144$. Thus, in the range $a\in (0.0144,0.0167)$ the region of stability consists of two disconnected parts separated by a region, where the circular orbits become unstable. For every spin parameter $a\in (0.0144,0.0167)$ the region of instability is delimited by two marginally stable circular orbits located at radii, which correspond to the solutions of the equation $\omega_r^2=0$.  Increasing the angular momentum the region of instability becomes smaller while at $a=0.0167$ it vanishes completely.

These configurations have astrophysical implications since they lead for example to a discontinuity in the accretion disk within the thin disk model. In this case the accretion disk consists of two annular regions separated by a gap. Such behavior is not uncommon among the various compact objects, and it arises for example in some naked singularity spacetimes like the Janis-Newman-Winicour solution.

For the counter-rotating orbits as well as for the static wormhole solution the region of stability resembles the case of the Kerr black hole. There is an innermost stable circular orbit (ISCO) located at the radial distance $r_{ISCO}$, where the radial epicyclic frequency vanishes, and all the orbits at higher values of the radial coordinate are stable. The ISCO takes its closest position to the wormhole throat at the static limit located at $r/r_0= 2$. When the angular momentum of the wormhole increases, it moves away to larger radii.

\section{Properties of the epicyclic frequencies}

The epicyclic and orbital frequencies are the main quantities, which are used in developing geodesic models for the quasi-periodic oscillations from the accretion disk such as the orbital precession and resonant models. Hence, their properties determine important characteristics of the model like the possible types of resonances, which can be excited, the radial distance from the compact object, where the resonance process takes place, and the values of the observable peak frequencies.

\begin{figure}[t!]
    		\setlength{\tabcolsep}{ 0 pt }{\small\tt
		\begin{tabular}{ cc}
           \includegraphics[width=0.5\textwidth]{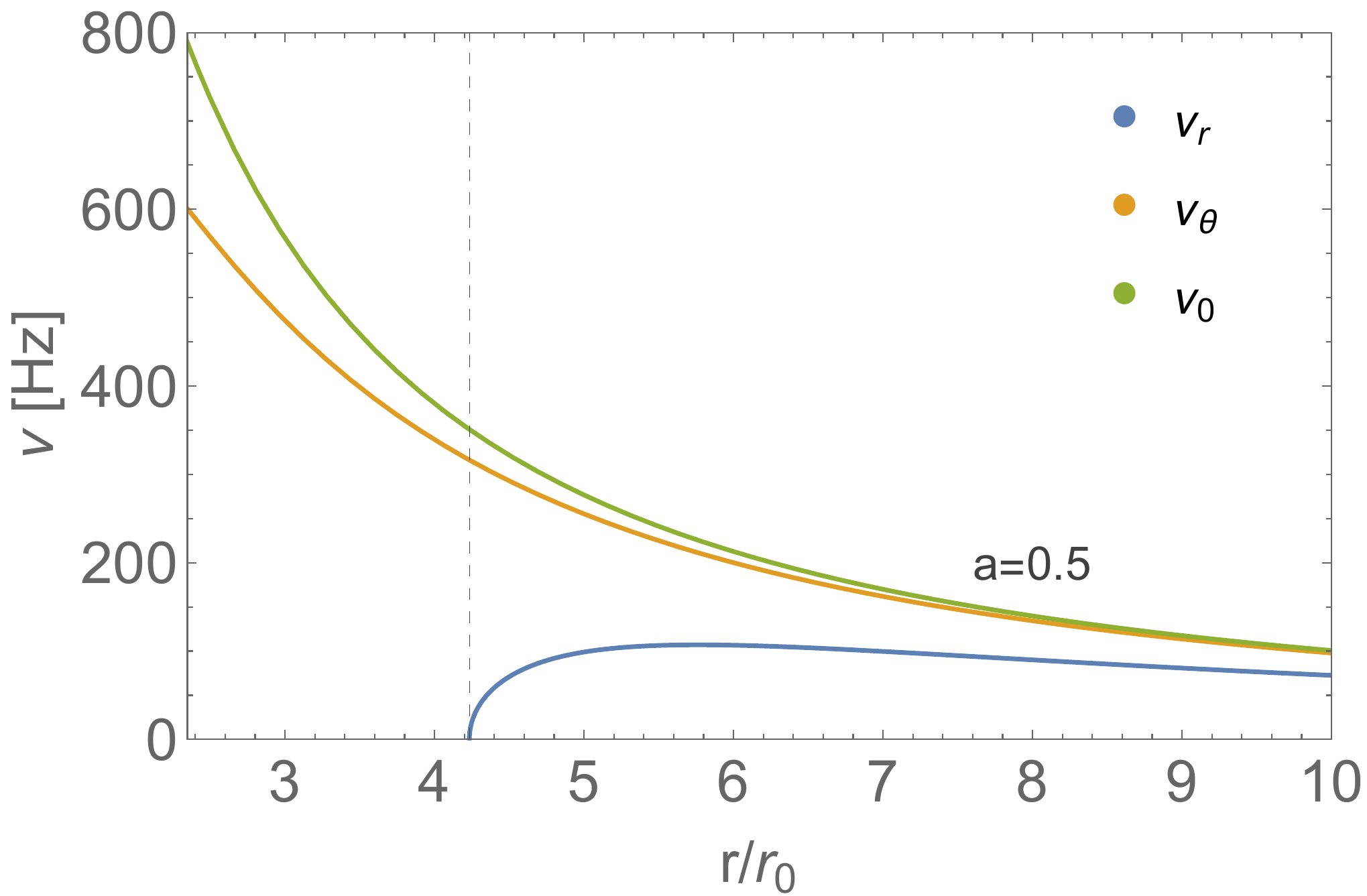}
		   \includegraphics[width=0.5\textwidth]{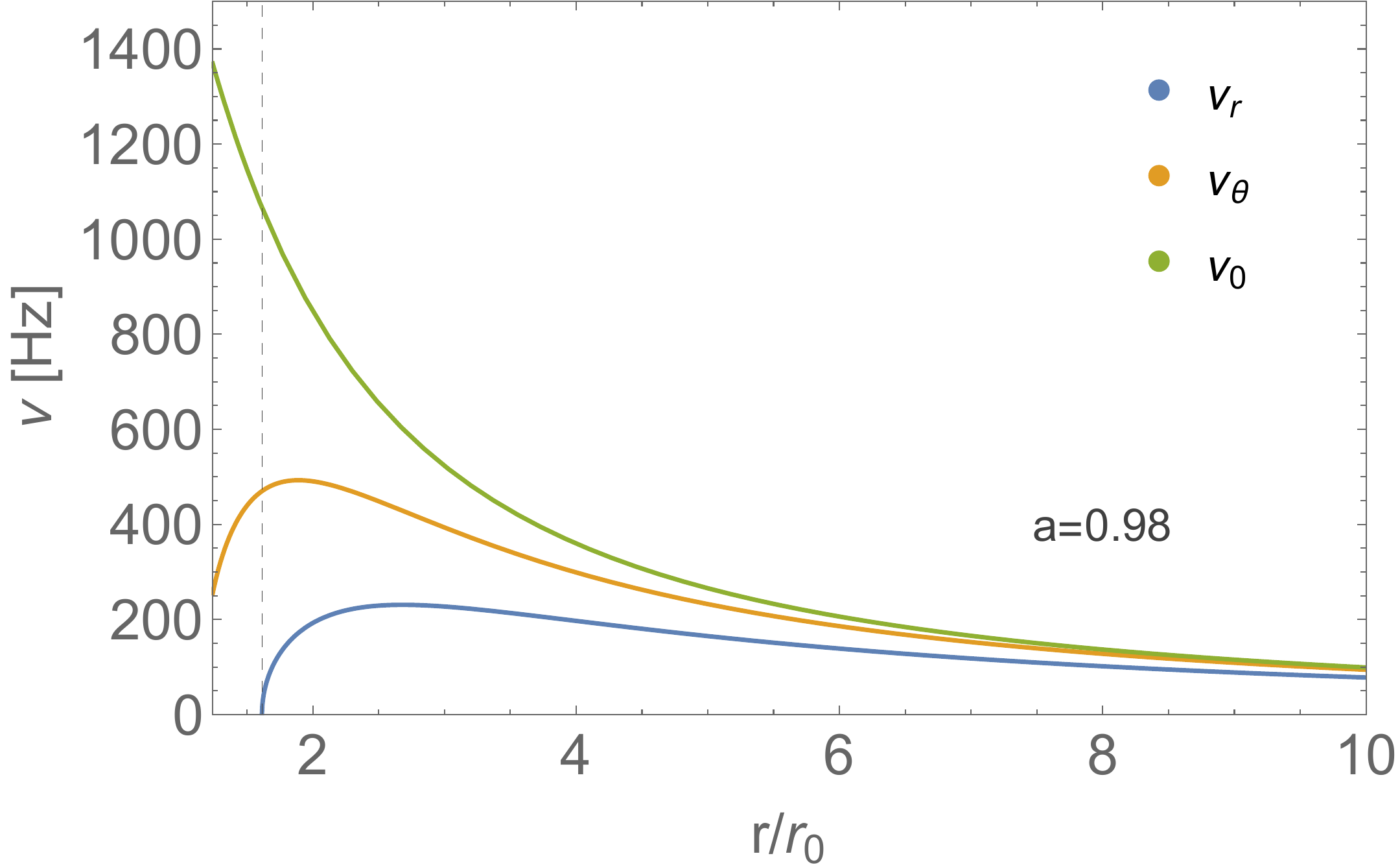} \\[1mm]
           \hspace{1.2cm}  $a)$ \hspace{7cm}  $b)$ 
     \end{tabular}}
 \caption{\label{fig:frequency_Kerr}\small Examples of the qualitatively different types of behaviour of the epicyclic frequencies for the Kerr black hole. For slow rotation the vertical epicyclic frequency is a monotonically decreasing function, while for higher spins it possesses a single maximum. The radial coordinate takes values larger than the photon orbit, and we represent the location of the ISCO with a dashed line. }
\end{figure}

In this section we will examine the wormhole epicyclic frequencies making comparison with the Kerr black hole. For the Kerr black hole we observe only a slight variation in the behavior of the characteristic frequencies in the whole range of the spin parameter $a\in [0,1]$. The orbital frequency is a  monotonically decreasing function for any value of the spin parameter and the radial coordinate above the photon orbit. The radial epicyclic frequency always possesses a single maximum, while the vertical one is a monotonically decreasing function for slow rotation and gets a single maximum for rapidly rotating black holes. In addition, for any value of the spin parameter the orbital frequency is always larger than the vertical epicyclic frequency, which on the other hand is larger than the radial one. Thus, we have the ordering $\omega^2_0>\omega^2_\theta > \omega^2_r$ for the whole range of the radial coordinate above the photon orbit\footnote{We should note that violating the Kerr limit for the spin parameter $a\leq 1$ changes  the properties of the epicyclic frequencies, and for naked singularities different situations can exist \cite{Torok:2005b}.}. The behavior of the  frequencies for the Kerr black hole is demonstrated in fig. $\ref{fig:frequency_Kerr}$ for some characteristic values of the spin parameter for the two qualitatively different cases.

\begin{figure}[t!]
            \centering
    		\setlength{\tabcolsep}{ 0 pt }{\small\tt
		    \includegraphics[width=0.6\textwidth]{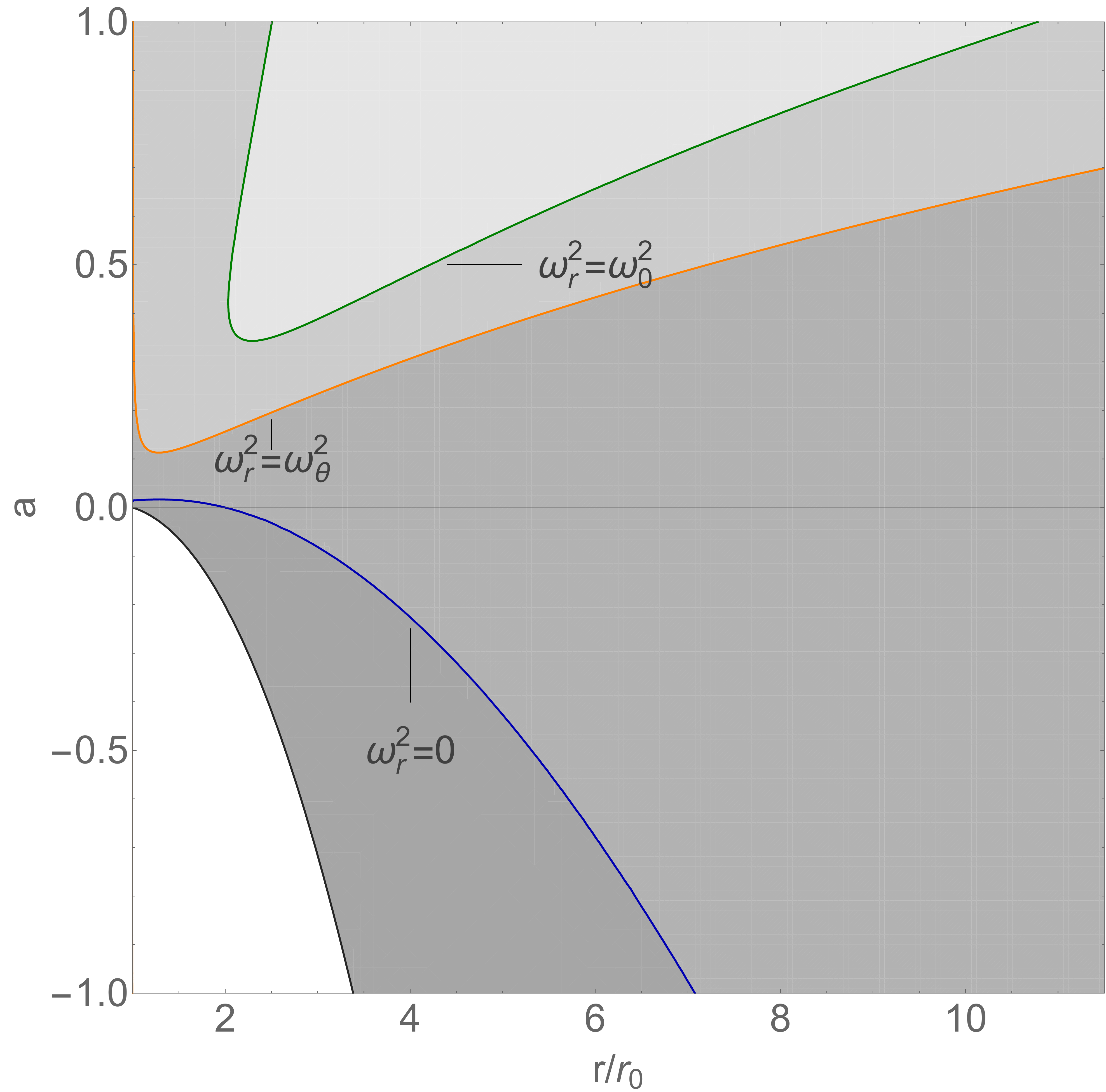}}		
 \caption{\label{fig:omega_xy}\small Ordering of the orbital and the epicyclic  frequencies for rotating traversable wormholes.  The curve $\omega^2_r = \omega^2_\theta$ is plotted in orange, while the curve $\omega^2_r = \omega^2_0$ is represented in green. In the region above the orange curve we have $\omega^2_r > \omega^2_\theta$, and above and bounded by the green curve $\omega^2_r > \omega^2_0$ is satisfied. For counter-rotating orbits with $a<0$ it is fulfilled that  $\omega^2_\theta > \omega^2_0$, while in the co-rotating case $a>0$ we have the opposite inequality. The grey and the blue curves are the boundaries of the regions of existence and stability of the timelike circular orbits, respectively. }
\end{figure}

While sharing some similarities with the Kerr black hole, the epicyclic frequencies for the wormholes show also major distinctions. One of the important differences is that various orderings of the orbital and epicyclic frequencies can be realized. This enables much more diverse scenarios of resonance excitation, some of which leading to stronger observable signals. In fig. $\ref{fig:omega_xy}$ we illustrate the possible cases for the particular wormhole solution ($\ref{wormhole}$)  by plotting the curves $\omega^2_r = \omega^2_\theta$ and $\omega^2_r = \omega^2_0$. In the region above and bounded by each of the curves the inequalities $\omega^2_r > \omega^2_\theta$ and $\omega^2_r > \omega^2_0$ are satisfied, respectively. From the expressions for the epicyclic frequencies ($\ref{omega_xy}$) we can see that for co-rotating orbits we have $\omega^2_\theta < \omega^2_0$, while for counter-rotating orbits the opposite case $\omega^2_\theta > \omega^2_0$ is realized. This result is rather general for the traversable wormholes, since it applies for the whole class of wormhole solutions with metric function $N$, $\omega$, $b$ and $K$ depending only on $r$.

The analysis performed in fig. $\ref{fig:omega_xy}$ shows that the case of counter-rotating orbits resembles the Kerr black hole since it possesses a uniform frequency ordering $\omega^2_\theta > \omega^2_0>\omega^2_r$ for the whole range of the radial coordinate $r\in(r_{ISCO}, \infty)$. For co-rotating orbits we have various scenarios. In the region above the curve $\omega^2_r = \omega^2_0$ we have $\omega^2_r > \omega^2_0 > \omega^2_\theta$ , below the curve $\omega^2_r = \omega^2_\theta$ the inequality $\omega^2_0>\omega^2_\theta >\omega^2_r$ is satisfied, while  between the two curves we have $\omega^2_0>\omega^2_r>\omega^2_\theta$. In the static limit we get the degenerate case $\omega^2_\theta = \omega^2_0>\omega^2_r$ similar to the Schwarzschild black hole. In the next section we will examine the implications of these types of frequency ordering on the possibilities for formation of different resonances.

Next we study the behavior of the orbital and epicyclic   frequencies as a function of the radial coordinate $r$ for various spin parameters $a$. For counter-rotating orbits we observe again a  consistent behavior for the whole range of $a \in (0,1]$, while for co-rotating orbits we have different possibilities depending on the speed of rotation of the wormhole. In the counter-rotating case the radial epicyclic frequency always has a single maximum, while the vertical one is a monotonically decreasing function of $r$. Thus, for counter-rotating orbits the wormhole spacetime resembles the slowly rotating Kerr black hole. In the co-rotating case we can classify the wormhole solution with respect to the behavior of the epicyclic frequencies in the following categories:

\begin{description}
\item I. $a\in[0,0.0144)$: The region of stability of the circular orbits is simply connected. The radial epicyclic frequency $\omega_r$ has a single maximum, while $\omega_\theta$ is a monotonically decreasing function.
\item II. $a\in[0.0144, 0.0167]$: The region of stability of the circular orbits consists of two disconnected parts. In each of the regions of stable orbits the radial epicyclic frequency $\omega_r$ possesses a single maximum. The vertical epicyclic frequency $\omega_\theta$ is a monotonically decreasing function.
\item III. $a\in[0.0167, 0.025]$: The region of stability of the circular orbits is simply connected. The radial epicyclic frequency $\omega_r$ has  two maxima and a minimum, while $\omega_\theta$ is a monotonically decreasing function.
\item IV. $a\in[0.025, 0.029)$: The region of stability of the circular orbits is simply connected. The radial epicyclic frequency $\omega_r$ has two maxima and a minimum, while $\omega_\theta$ possesses a single maximum.
\item V. $a\in[0.029, 1]$: The region of stability of the circular orbits is simply connected. Both the radial and the vertical epicyclic frequencies possess a single maximum.
\end{description}

\begin{figure}
    		\setlength{\tabcolsep}{ 0 pt }{\small\tt
		\begin{tabular}{ cc}
           \includegraphics[width=0.5\textwidth]{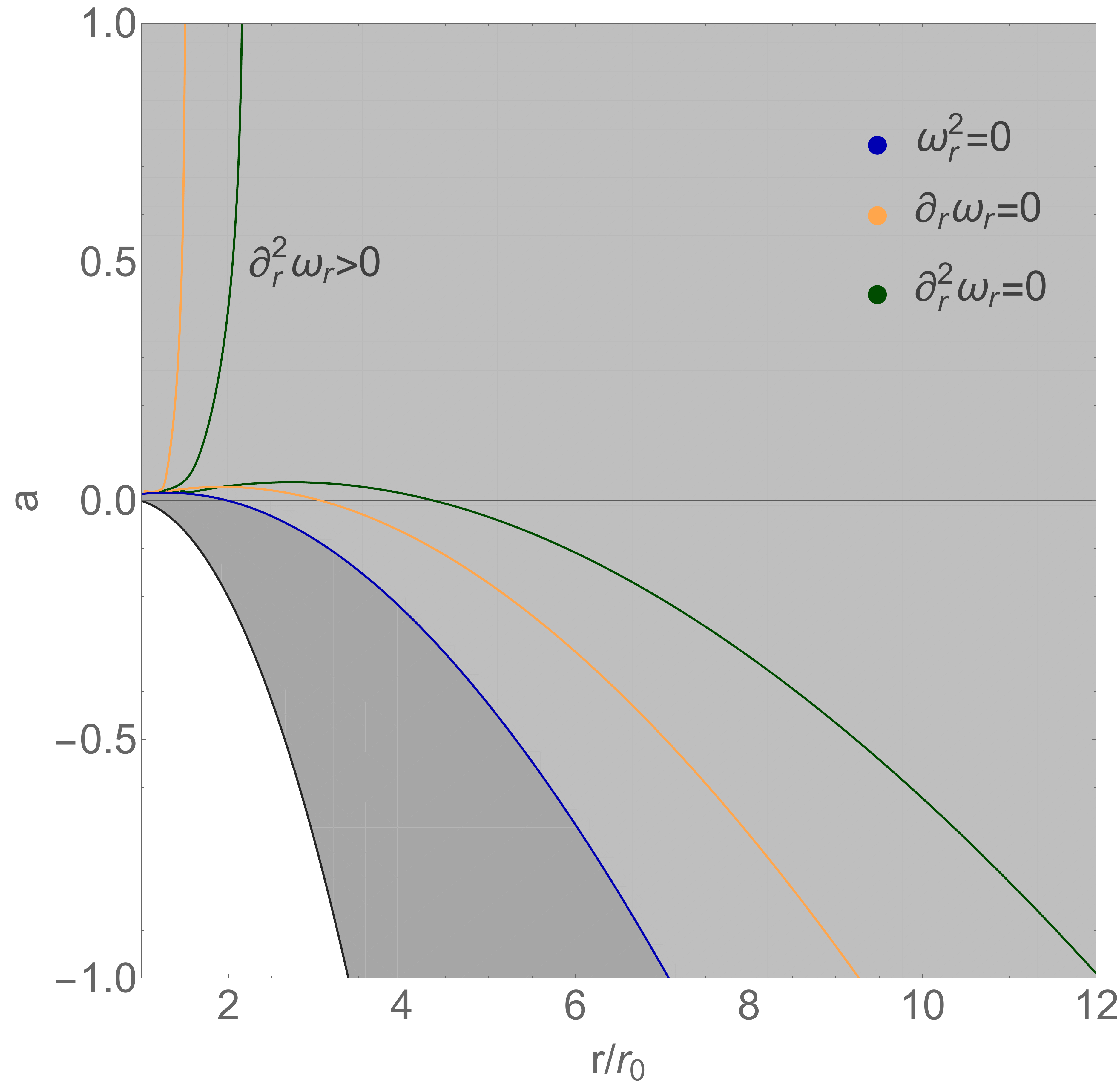}
		   \includegraphics[width=0.5\textwidth]{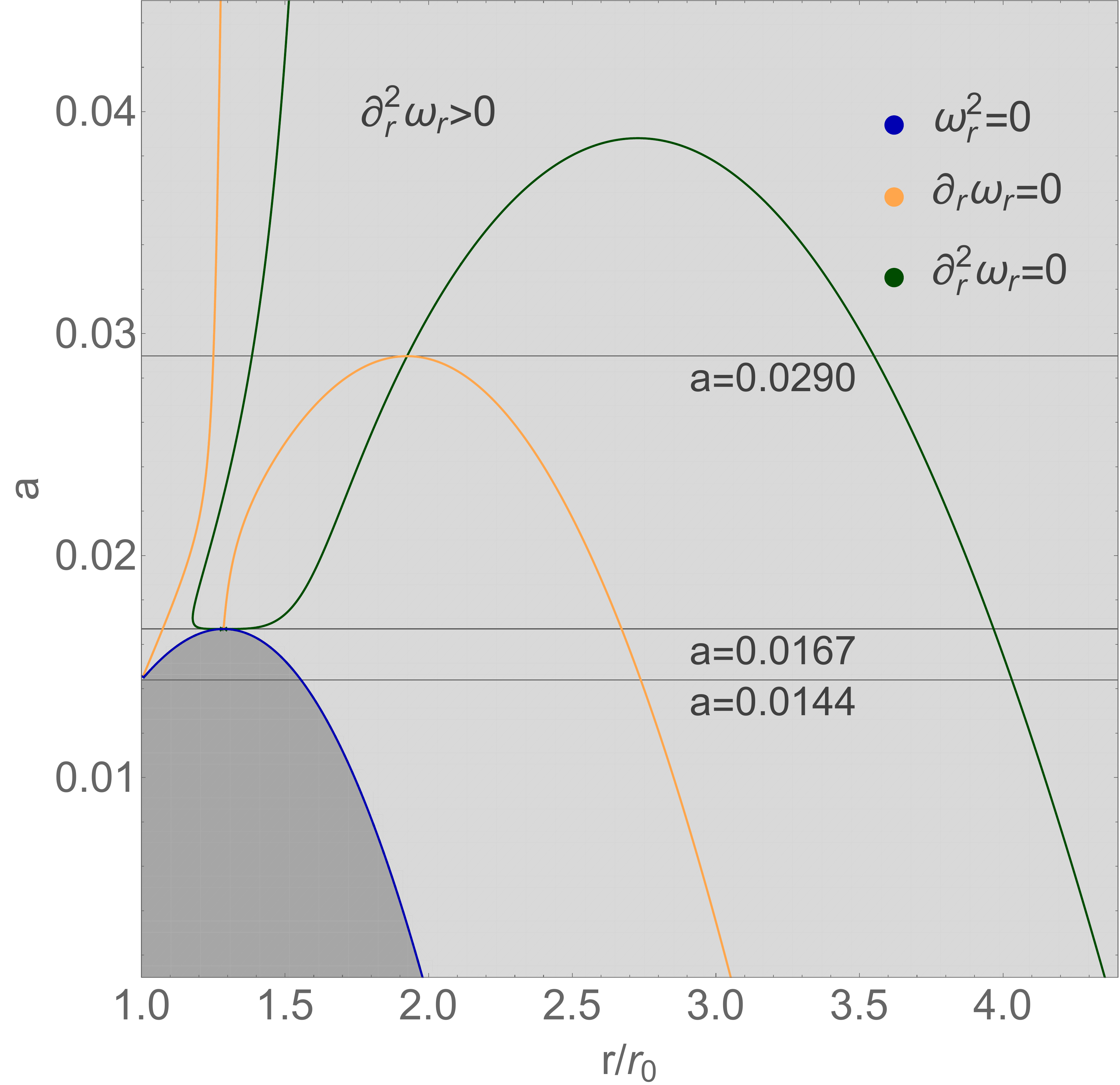} \\[1mm]
           \hspace{1.2cm}  $a)$ \hspace{7cm}  $b)$ \\[3mm]
           \includegraphics[width=0.5\textwidth]{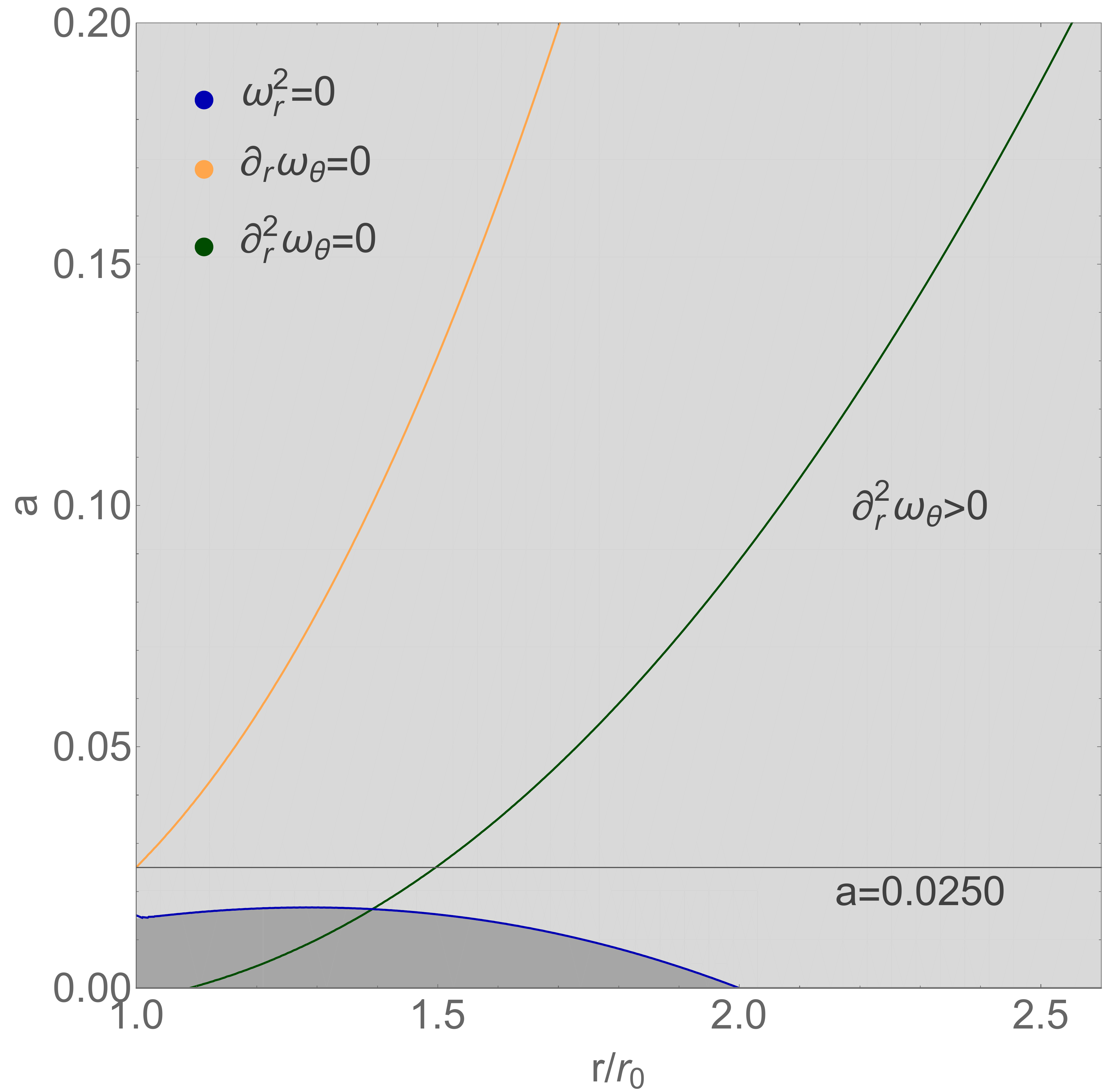} \\[1mm]
           \hspace{1.2cm}  $c)$
        \end{tabular}}
 \caption{\label{fig:frequency_max}\small Behavior of the radial epicyclic frequency a) and b), and the vertical epicyclic frequency c) as a function of the radial coordinate. The curves $\partial_r\omega_r =0$ and $\partial_r\omega_\theta =0$ are represented in orange, while the second derivatives $\partial^2_r\omega_r =0$ and $\partial^2_r\omega_\theta =0$ are represented in green. The functions $\partial^2_r\omega_r$ and $\partial^2_r\omega_\theta$ are positive above and on the righthand side of the green curves.  In the zoomed plot b) we illustrate the regions with different types of behavior of the radial epicyclic frequency for slow rotation, where we denote the transition values of the spin parameter with horizontal lines. We further show the boundaries of the domain of existence and stability of the circular geodesics with grey and blue lines, respectively. }
\end{figure}

We see that the epicyclic frequencies for the counter-rotating case and the very slowly co-rotating case I. behave like the slowly rotating Kerr black hole, as the static limit resembles the Schwarzschild black hole. Then, we have some exotic regions in the parametric space II., III. and IV. with a multi-connected region of stability of the circular orbits or multiple extrema of the radial epicyclic frequency, which don't exist for the Kerr black hole. Increasing further the spin parameter in region V. the epicyclic frequencies start to behave like for the rapidly rotating Kerr black hole. The orbital frequency is always a monotonically decreasing function of the radial coordinate both for co- and counter-rotating orbits.

\begin{figure}
    		\setlength{\tabcolsep}{ 0 pt }{\small\tt
		\begin{tabular}{ cc}
           \includegraphics[width=0.5\textwidth]{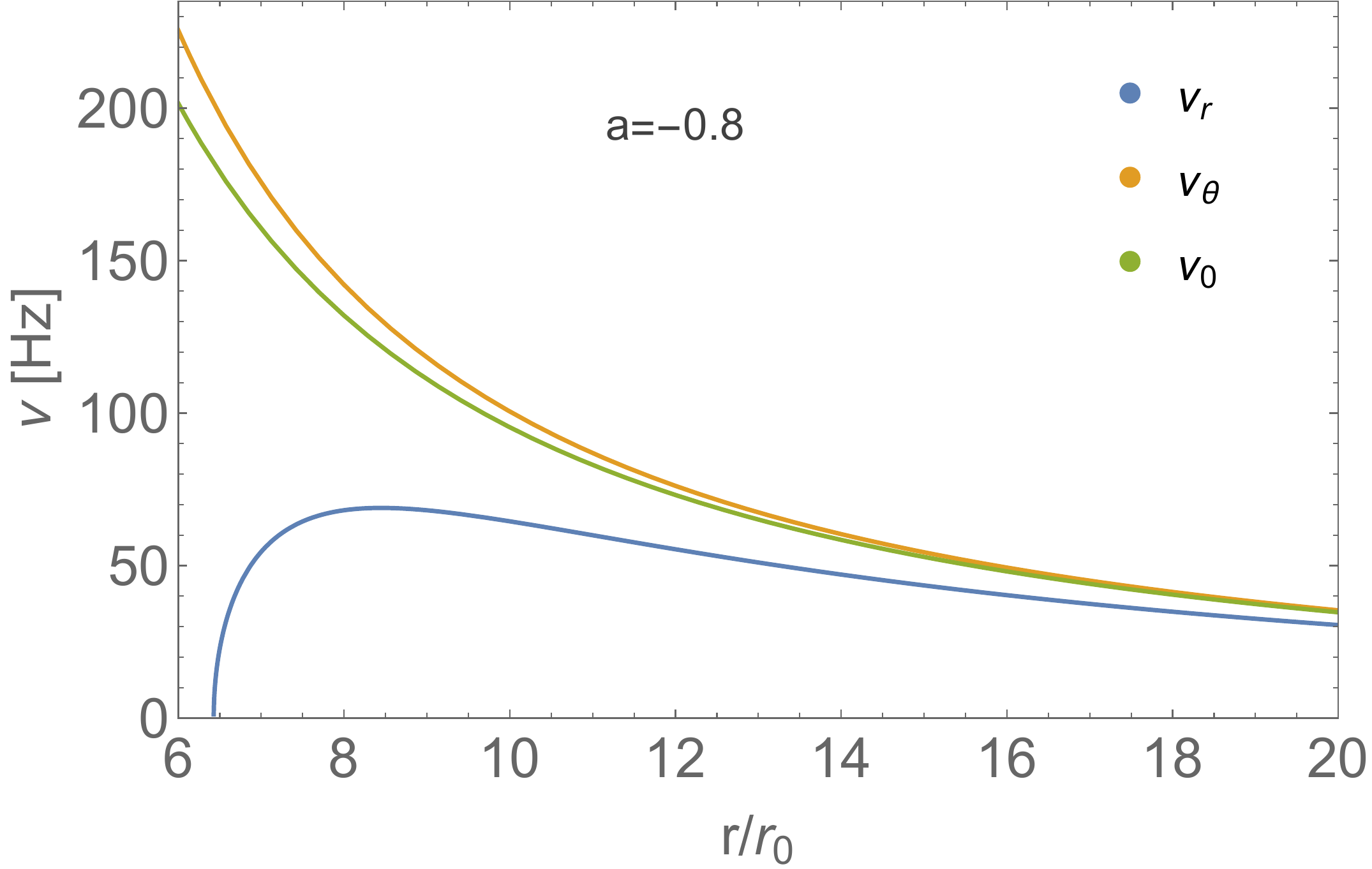}
           \includegraphics[width=0.5\textwidth]{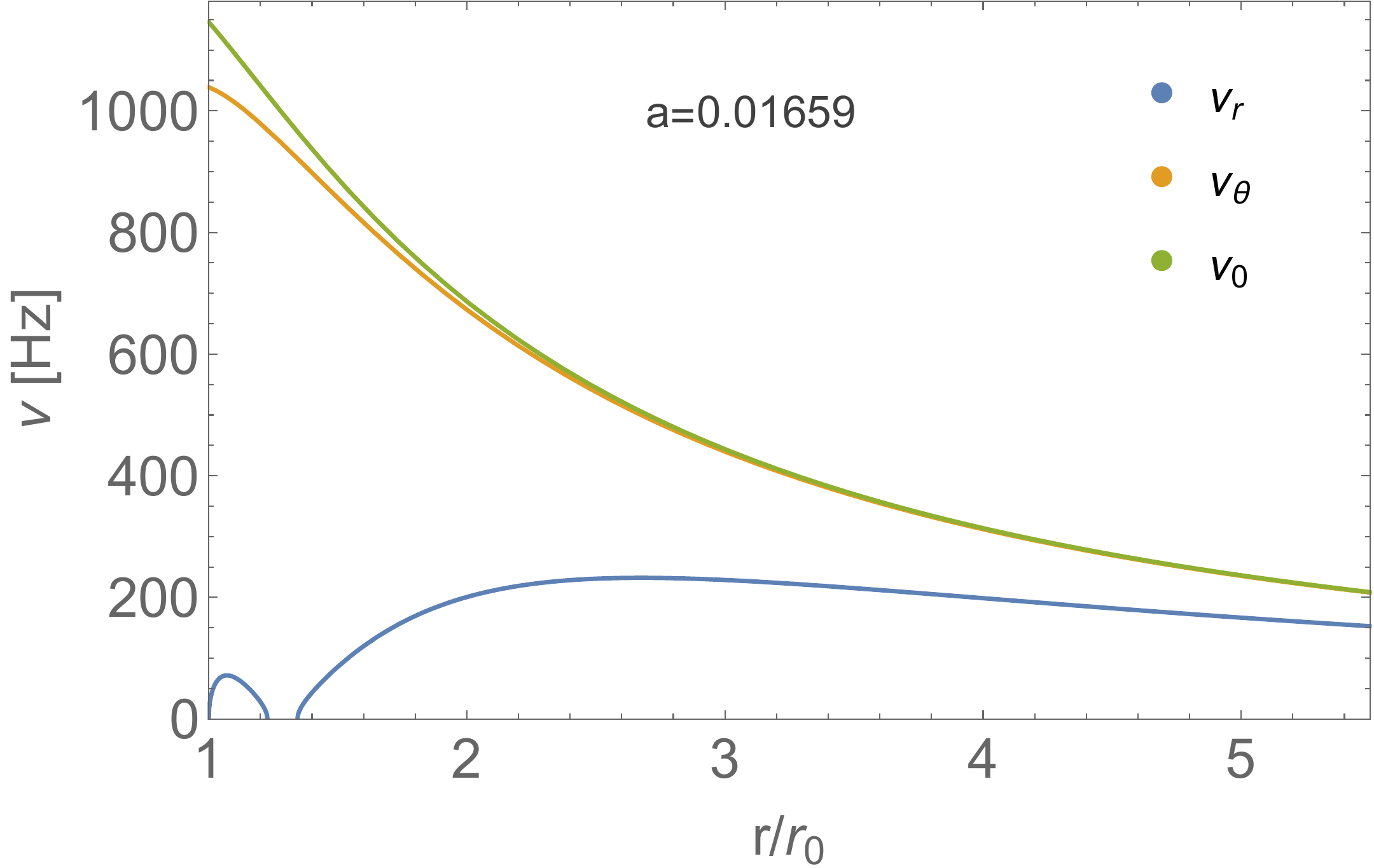}\\[1mm]
           \hspace{1.2cm}  $a)$ \hspace{7cm}  $b)$ \\[3mm]
            \includegraphics[width=0.5\textwidth]{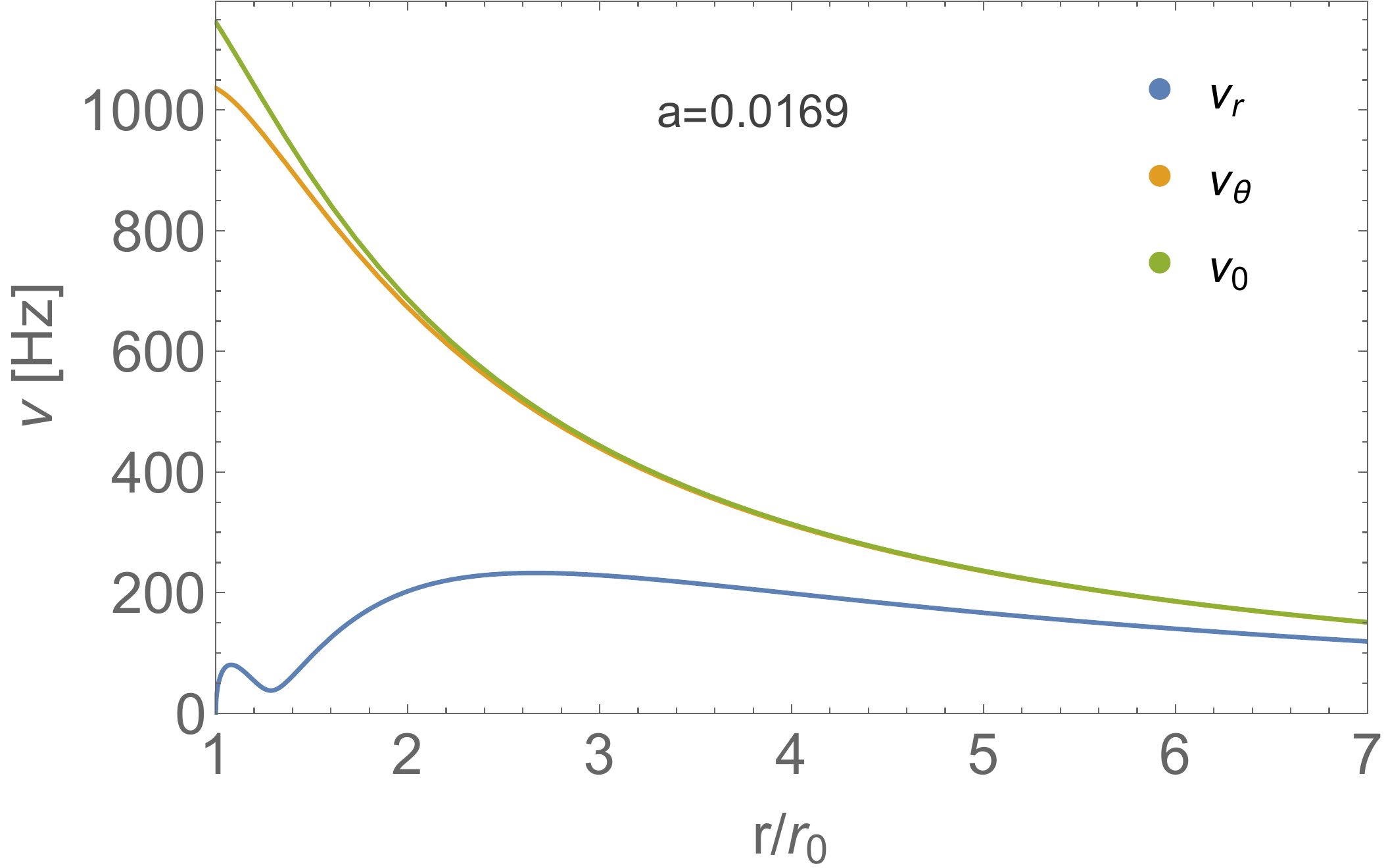}
           \includegraphics[width=0.5\textwidth]{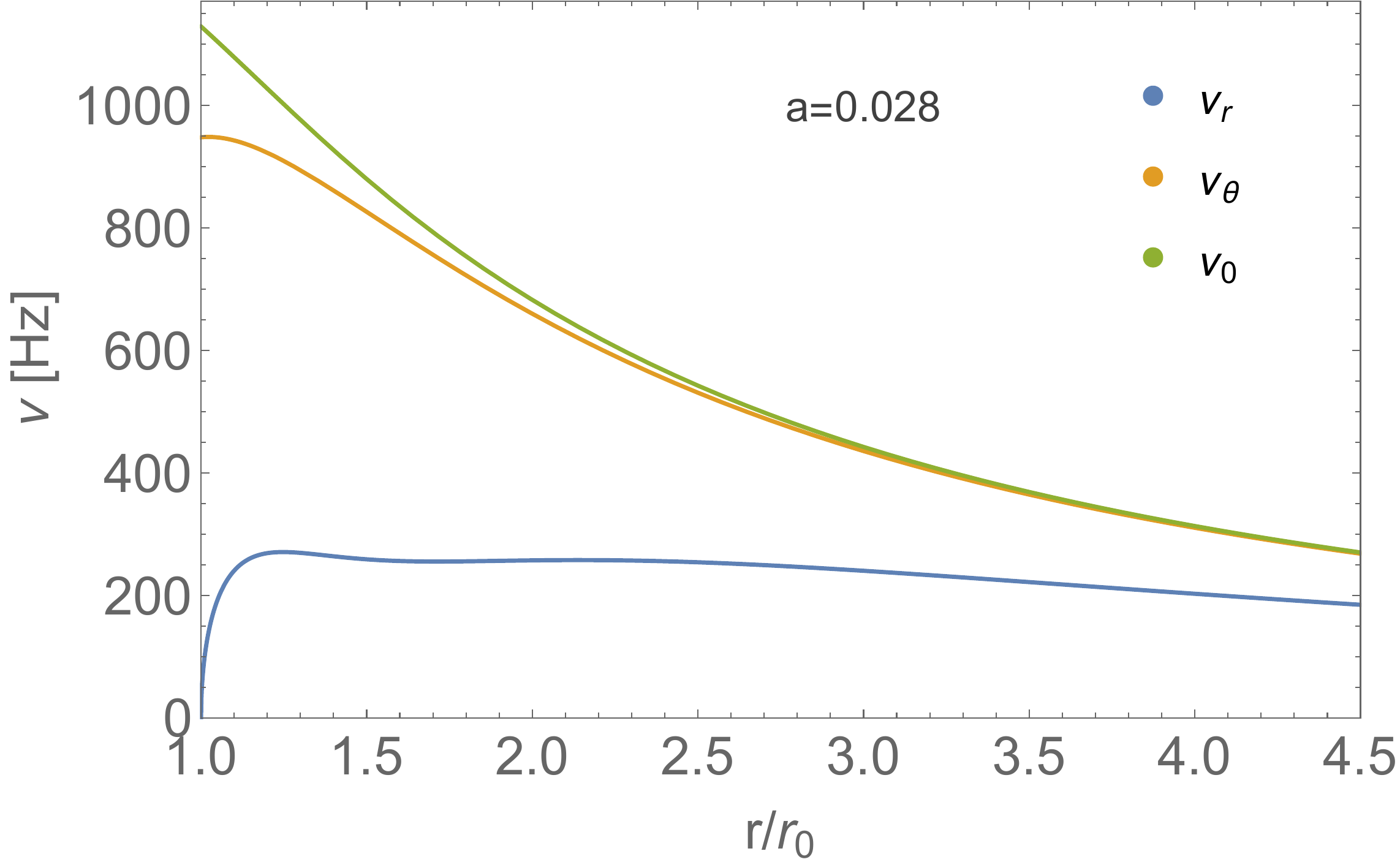}\\[1mm]
           \hspace{1.2cm}  $c)$ \hspace{7cm}  $d)$ \\[3mm]
           \includegraphics[width=0.5\textwidth]{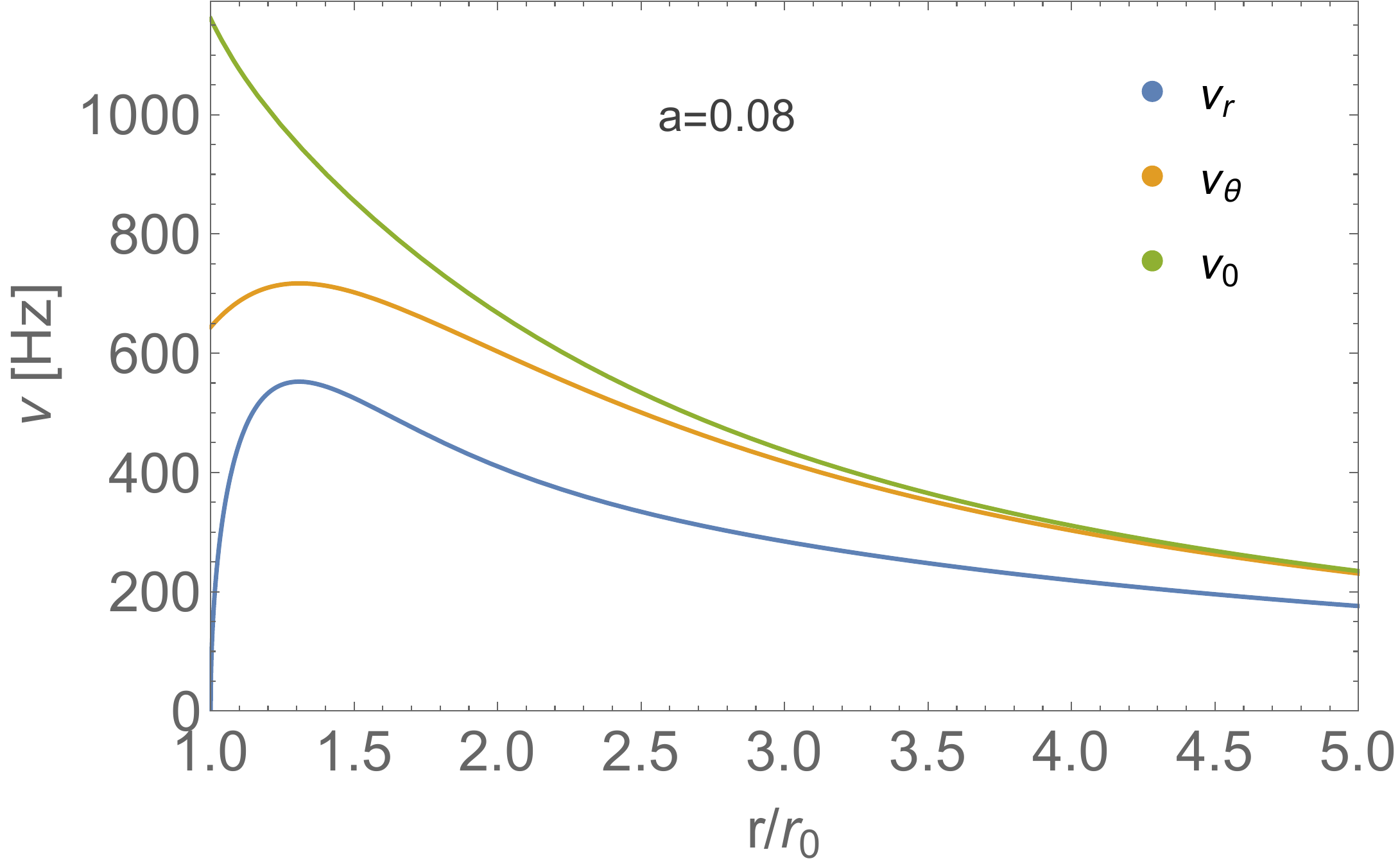}\\[1mm]
            \hspace{1.2cm}           $e)$
        \end{tabular}}
 \caption{\label{fig:frequency}\small Examples of the qualitatively different types of behaviour of the epicyclic frequencies for rotating traversable wormholes.  }
\end{figure}

In fig. $\ref{fig:frequency_max}$ we present the analysis of the behavior of the epicyclic frequencies as a function of $r$ for different values of the spin parameter by plotting the curves $\partial_r\omega_r =0$ and $\partial_r\omega_\theta=0$, as well as the second derivatives $\partial^2_r\omega_r =0$, and $\partial^2_r\omega_\theta =0$. The regions of  the qualitatively different types of behavior are limited by horizontal lines corresponding to the characteristic values of the spin parameter, where the transitions occur. We further demonstrate examples of each of the classes I. - IV. in fig. $\ref{fig:frequency}$, where we plot the frequencies $\nu_r = \omega_r/2\pi$, $\nu_\theta = \omega_\theta/2\pi$ and $\nu_0 = \omega_0/2\pi$ for some particular values of the spin parameter.

\section{Non-linear resonances}

In linear approximation the small deviations from circular geodesic motion are described by two independent harmonic oscillations with eigenfrequencies called radial and vertical epicyclic frequencies. However, a more realistic description of the processes in the accretion disk requires to include further non-linear terms in the perturbation equations. They give account for different forces exerted in the accreting fluid such as pressure, viscosity, magnetic fields etc., and lead to the coupling of the two epicyclic modes. Typically such interaction between the eigenfrequencies is a prerequisite for the excitation of resonances in dynamical systems, which are realized when the system reaches suitable conditions. Indeed, analytical and numerical investigations of different models of accretion show that resonances are frequently  present in the accretion disks and seem to be their intrinsic feature \cite{Rebusco}-\cite{Horak:2004b}.

Currently the physical processes which take part in the accretion disk are not understood sufficiently well. Therefore, it is difficult to derive rigorous expressions for the non-linear terms governing the behavior of the small perturbations. A reasonable approach is to consider some basic types of interactions, which are generic enough to arise in many physical situations for a wide range of particular processes. For example, we can consider  non-linear corrections to the perturbation equations describing the small deviation from geodesic circular motion in the form

\begin{eqnarray}
&&\frac{d^2 \xi^r}{dt^2} + \omega_r^2 \xi^r =  \omega_r^2 f_r\left(\xi^r, \xi^\theta, \frac{d\xi^r}{dt}, \frac{d\xi^\theta}{dt}\right), \nonumber \\[2mm]
&&\frac{d^2 \xi^\theta}{dt^2} + \omega_\theta^2 \xi^\theta =  \omega_\theta^2 f_\theta\left(\xi^r, \xi^\theta, \frac{d\xi^r}{dt}, \frac{d\xi^\theta}{dt}\right) ,
\end{eqnarray}
where $f_r$ and $f_\theta$ are non-linear functions. The specific form of these functions should be determined by the properties of the physical model of the accretion flow. However, without resorting to a particular model, we can suggest some simple cases, which are likely to arise in many scenarios and investigate their behavior. One of the simplest situations is to assume that $f_r=0$ and  $f_\theta = h\,\xi^r\xi^\theta$, where $h$ is a coupling constant. Then, the equation for the vertical oscillations takes the form

\begin{eqnarray}
\frac{d^2 \xi^\theta}{dt^2} + \omega_\theta^2 \xi^\theta = - \omega_\theta^2 h \cos(\omega_r t) \xi^\theta.
\end{eqnarray}
In this way we obtain the Mathieu equation, which is known to describe parametric resonances for ratios of the frequencies

\begin{eqnarray}
\frac{\omega_r}{\omega_\theta}= \frac{2}{n},
\end{eqnarray}
where $n$ is a positive integer (see e.g. \cite{Landau}). When the coupling is weak, or $h<<1$, the smallest possible value of $n$ corresponds to the strongest resonance. Despite that parametric resonances were obtained by adopting an ansatz for the frequencies coupling it was demonstrated that they are a mathematical property of thin, nearly Keplerian disks \cite{Abramowicz:2003}, \cite{Rebusco}, \cite{Horak:2004b}.

Another common dynamical system which exhibits resonant behavior is the forced non-linear oscillator. In this respect it was suggested that the non-linear effects in the perturbations of the circular orbits can be described by including a periodic radial force in the equation for the vertical oscillations with frequency equal to the radial epicyclic frequency. Then, the equation for the vertical oscillations takes the form

\begin{eqnarray}
\frac{d^2 \xi^\theta}{dt^2} + \omega_\theta^2 \xi^\theta
+ [{\rm non \; linear \; terms \; in} \: \xi^\theta] &=&
h(r) \cos(\omega_r t).
\end{eqnarray}

Resonances are excited  when the epicyclic frequencies take integer ratios $\omega_\theta= n\omega_r$. Since the equation is non-linear, the resonant solution can contain also linear combinations of the epicyclic frequencies, which gives further possibilities for fitting the  frequencies of the quasiperiodic oscillations.

In the previous models for the non-linear effects in the oscillations of the circular orbits  resonances occurred due to the coupling of the two epicycylic frequencies. In general it is also possible to consider  interactions of one of the epicyclic frequencies and the orbital frequency leading to the so called Keplerian resonances. The Keplerian resonances are less motivated from a theoretical point of view since it is difficult to imagine realistic physical processes in the accretion disk which may cause their excitation. Nevertheless, there is no physical reason either, which prevents their existence, and they can be studied as a possible source for the quasi-periodic oscillations.

In the following discussion we will study how the described resonance phenomena can explain the observed  twin-peak  frequencies in the X-ray flux from accreting compact objects provided that the compact object is modelled by a rotating traversable wormhole. In the resonance models the twin-peak frequencies are explained by identifying them with suitable combinations of resonant frequencies so that the observational ratio between the lower ($\nu_L$) and the upper ($\nu_U$) frequencies is satisfied, i.e. $\nu_U:\nu_L= 3:2$.  In general, identifications with frequencies corresponding to lower order resonances are preferred, since they lead to larger amplitudes of the observed signal. For the parametric resonance this can be done directly by making the identifications $\nu_U = \nu_\theta = \omega_\theta/2\pi$ and $\nu_L= \nu_r= \omega_r/2\pi$. In the case of the Kerr black hole this is the lowest order parametric resonance since $n=1,2$ parametric resonances don't exist. If we consider the forced resonances, we need to identify the observed frequencies with linear combinations of the epicyclic frequencies in order to achieve the $3:2$ ratio. For the Kerr black hole the lowest order possible forced resonances are $n=2$ and $n=3$ when the epicyclic frequencies are related as  $\omega_\theta:\omega_r = 2:1$ and $\omega_\theta:\omega_r = 3:1$. Then, the observational ratio can be obtained by making the identifications $\nu_U = \nu_\theta + \nu_r =(\omega_\theta +\omega_r)/2\pi$ and  $\nu_L=\nu_\theta$, and $\nu_U = \nu_\theta$ and  $\nu_L=\nu_\theta -\nu_r= (\omega_\theta -\omega_r)/2\pi$, respectively. The simplest cases of the Keplerian resonances, which are possible in the spacetime of the Kerr black hole are $\omega_0:\omega_r = 3:2$, $\omega_0:\omega_r = 2:1$, or $\omega_0:\omega_r = 3:1$, and similarly if we consider coupling between the vertical epicyclic and the orbital frequencies.

\begin{figure}[t!]
    		\setlength{\tabcolsep}{ 0 pt }{\small\tt
		\begin{tabular}{ cc}
           \includegraphics[width=0.5\textwidth]{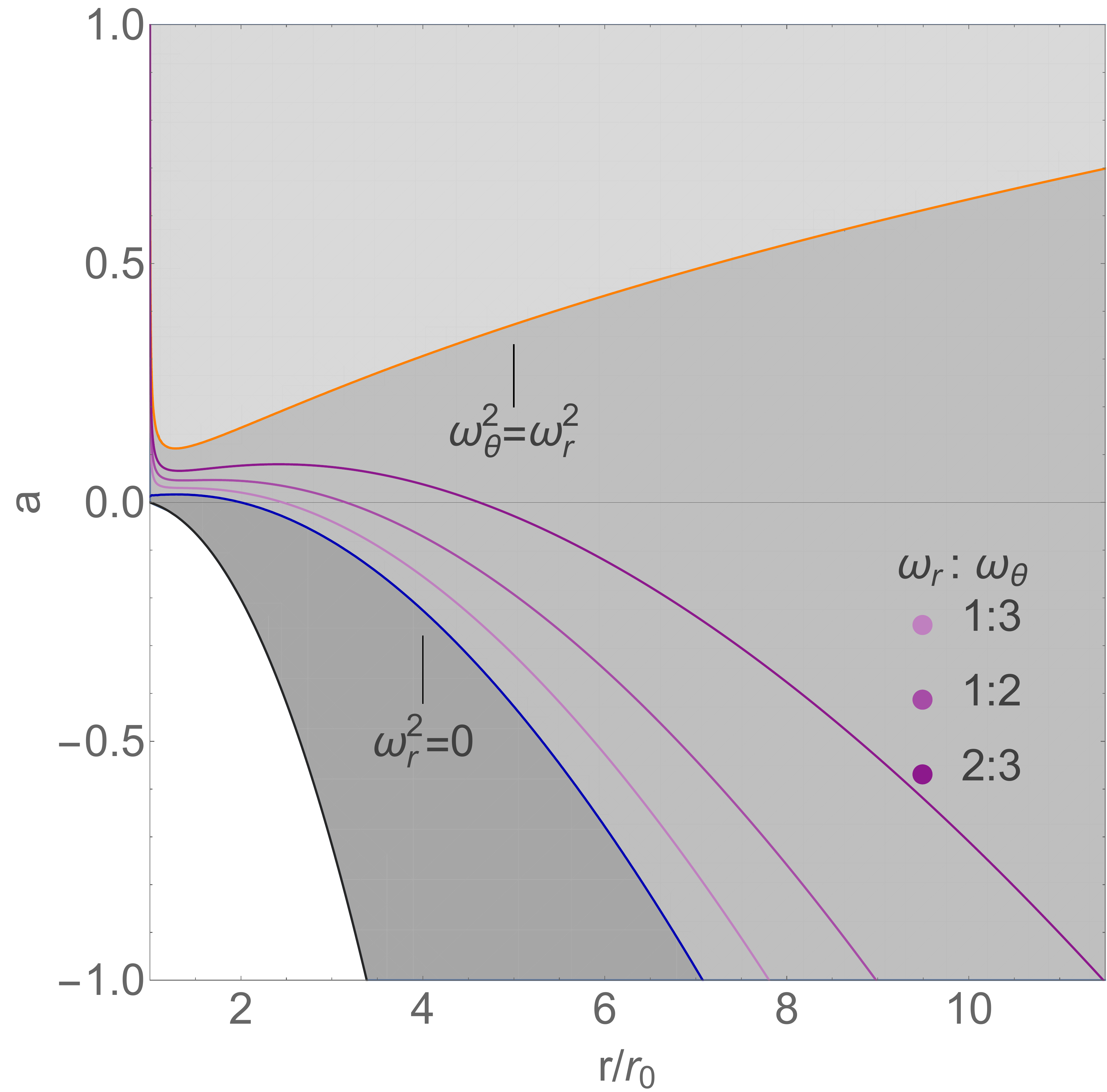}
		   \includegraphics[width=0.5\textwidth]{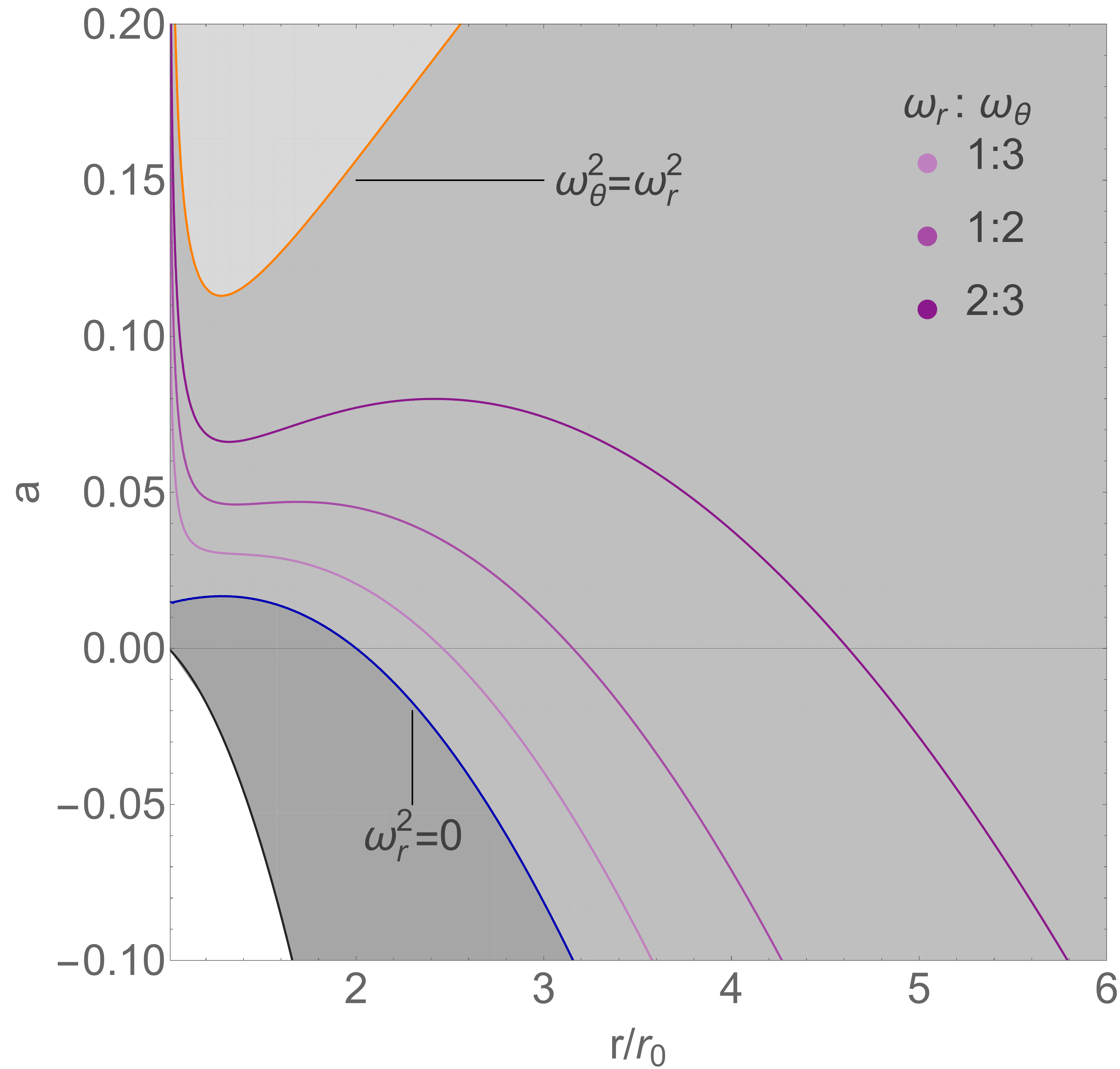} \\[1mm]
           \hspace{1.2cm}  $a)$ \hspace{7cm}  $b)$ \\[3mm]
           \includegraphics[width=0.5\textwidth]{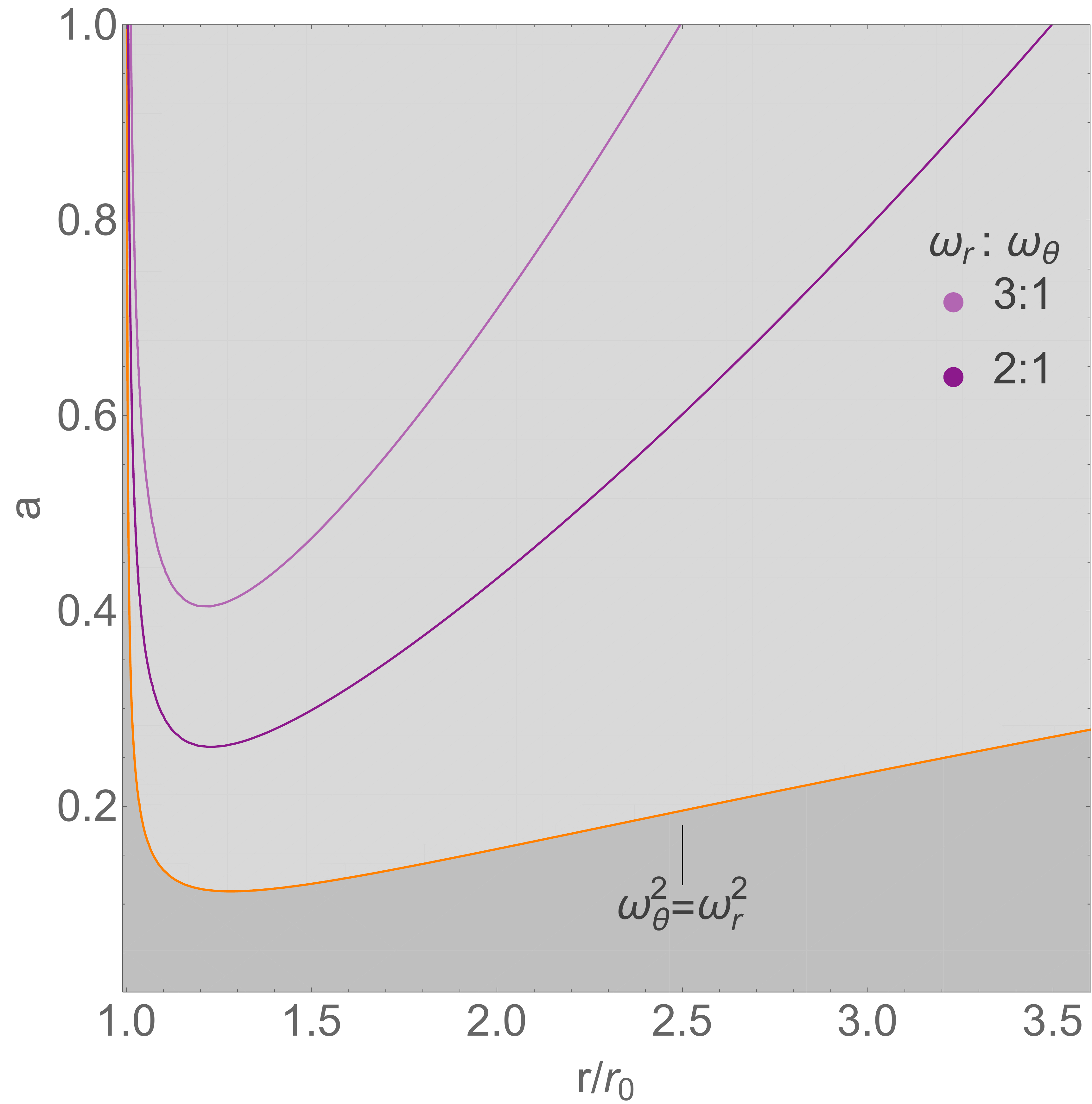} \\[1mm]
           \hspace{1.2cm}  $c)$
        \end{tabular}}
 \caption{\label{fig:res1}\small Location of the parametric and forced resonances depending on the wormhole spin. In a) and b) we  represent the case when the epicyclic frequencies satisfy the inequality $\omega_r < \omega_\theta$, while in c) we have the ordering $\omega_r > \omega_\theta$. In the zoomed plot b) we can see the location of the resonances for the static wormhole. }
\end{figure}

\begin{figure}
    		\setlength{\tabcolsep}{ 0 pt }{\small\tt
		\begin{tabular}{ cc}
           \includegraphics[width=0.5\textwidth]{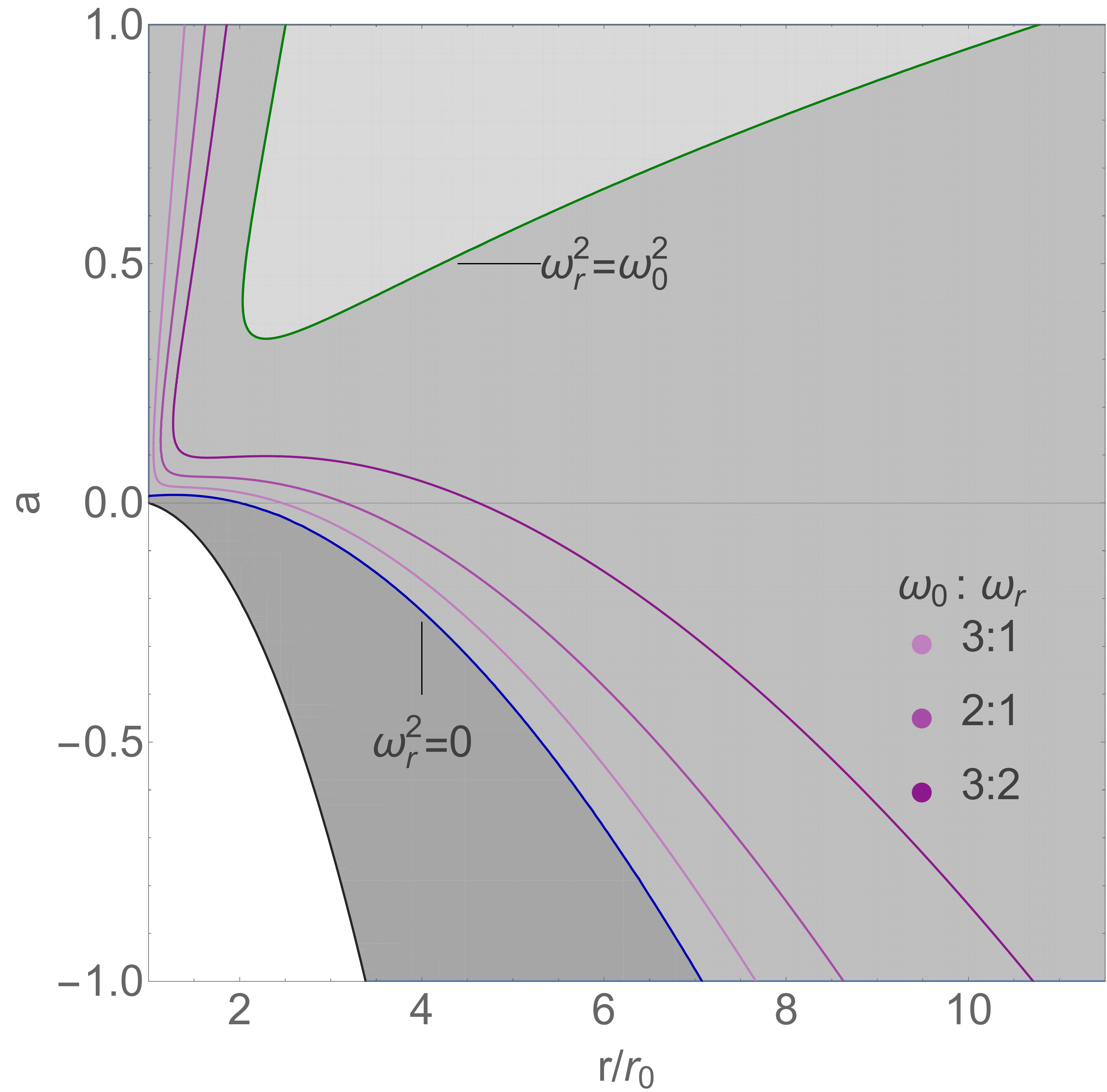}
		   \includegraphics[width=0.5\textwidth]{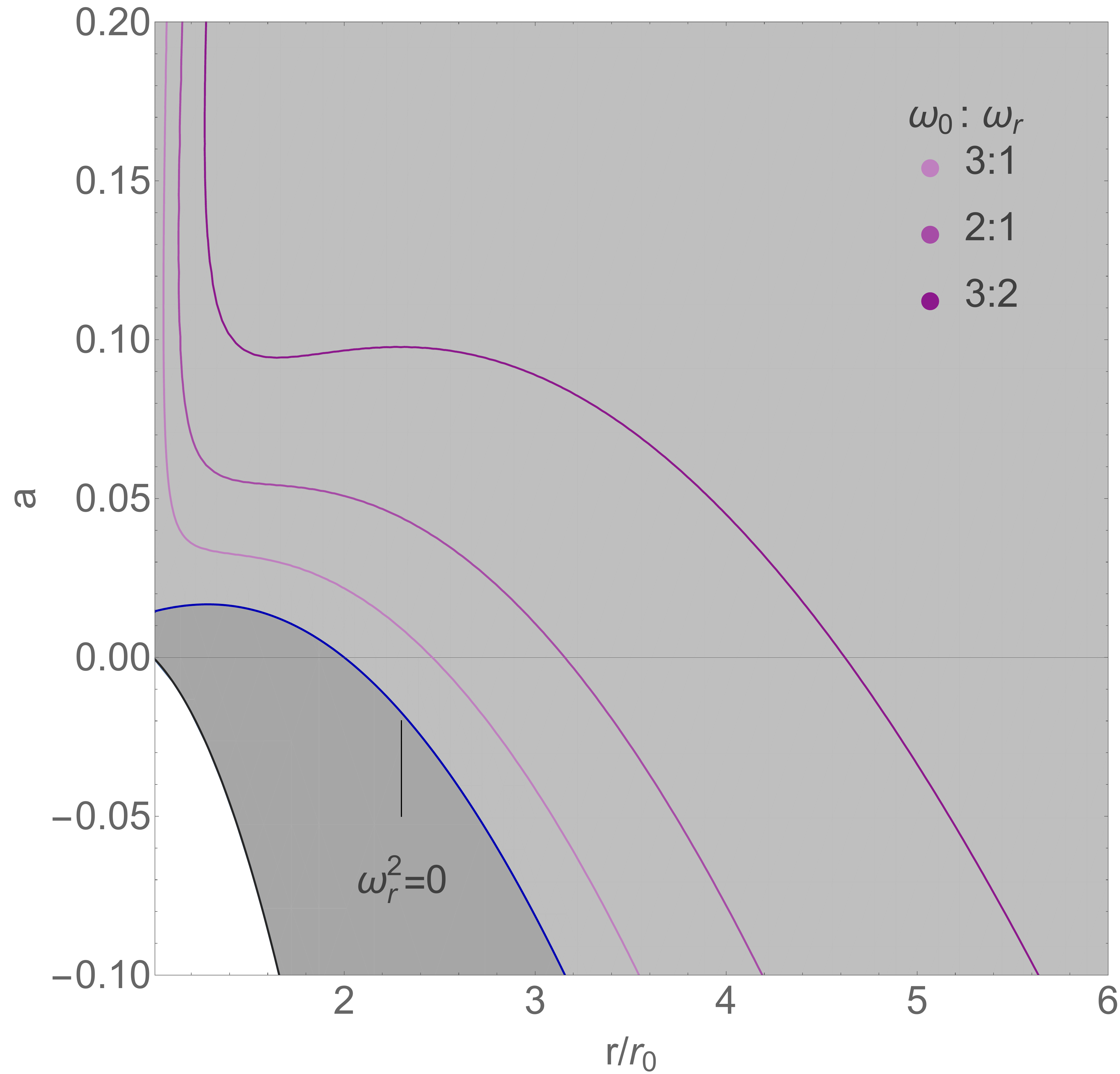} \\[1mm]
           \hspace{1.2cm}  $a)$ \hspace{7cm}  $b)$ \\[3mm]
           \includegraphics[width=0.5\textwidth]{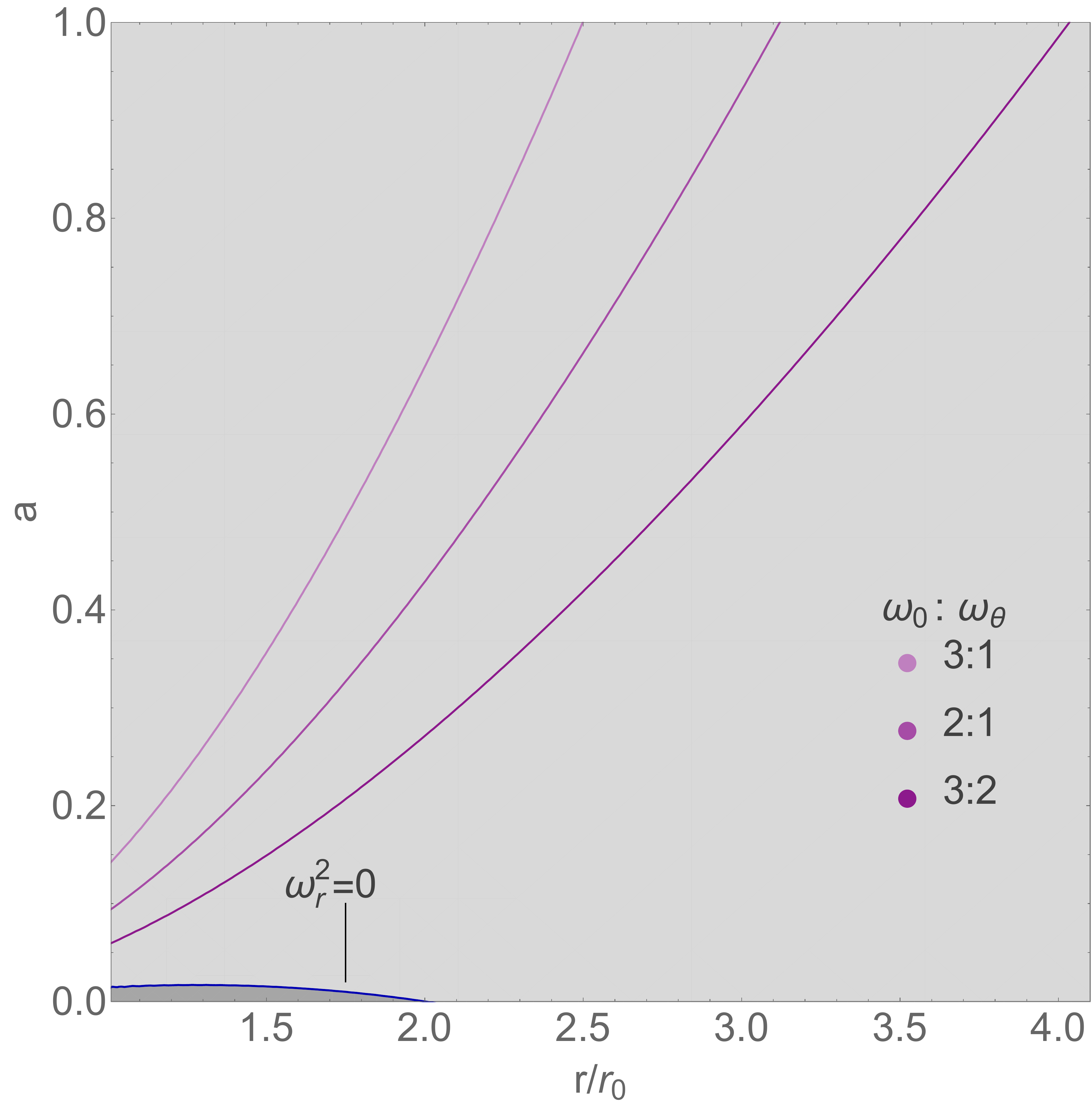} \\[1mm]
           \hspace{1.2cm}  $c)$
        \end{tabular}}
 \caption{\label{fig:res2}\small Location of the Keplerian resonances depending on the wormhole spin.  In a) and b) (zoomed plot) we study the resonances due to the coupling between the radial epicyclic and the orbital frequencies, while in c)  the coupling is between the vertical epicyclic and the orbital frequencies. The lowest order Keplerian resonances don't exist in the regions in the parametric space, where $\omega_0<\omega_r$ or $\omega_0<\omega_\theta$ is satisfied. }
\end{figure}

In wormhole spacetimes we obtain a much richer picture of possible resonant phenomena. One of the most distinctive features compared to the Kerr black hole is that various types of ordering of the orbital and epicyclic frequencies occur in the different regions of the parametric space. This allows for the excitation of more diverse types of resonances which don't exist in the Kerr spacetime. For the Kerr black hole we always have the inequality $\omega_\theta > \omega_r$. This prevents the excitation of the lower order parametric resonances $n=1$ and $n=2$, i.e. $\omega_r=2\omega_\theta$ and $\omega_r=\omega_\theta$, which would also possess the highest amplitudes. In contrast, for the rotating wormhole given by eq. ($\ref{wormhole}$) the $n=1$ and $n=2$ parametric resonances are possible for any value of the spin parameter $a\in [0,1]$. The observed $3:2$ ratio between the twin-peak frequencies can be explained by identifying the lower and upper observable frequencies $\nu_L$ and $\nu_U$ as $\nu_U = \nu_\theta + \nu_r$ and  $\nu_L=\nu_r$ in the $n=1$ case, and  $\nu_U = 3\nu_\theta =3\nu_r$ and $\nu_L=2\nu_r=2\nu_\theta$ in the $n=2$ case.

For the excitation of the lowest order forced resonances we obtain new possibilities when the epicyclic frequencies take the ratios $\omega_\theta:\omega_r = 1:2$ and $\omega_\theta:\omega_r = 1:3$. They result in the observable frequencies $\nu_U = \nu_\theta + \nu_r$,  $\nu_L=\nu_r$, and $\nu_U=\nu_r$, $\nu_L = \nu_r - \nu_\theta$, respectively. The Keplerian resonances can be excited for combinations like  $\omega_0:\omega_r= 3:2$ ($\nu_U = \nu_0$, $\nu_L = \nu_r$) , $\omega_0:\omega_r = 2:1$ ($\nu_U = 3\nu_r$, $\nu_L = \nu_0$), or $\omega_0:\omega_r= 3:1$ ($\nu_U = \nu_0$, $\nu_L = 2\nu_r$) in the regions where the ordering $\omega_0>\omega_r$ is valid, and the corresponding cases with coupling between the vertical epicyclic and the orbital frequencies  when $\omega_0>\omega_\theta$. Our investigations show that the lowest order Keplerian resonances with ratios between the orbital frequency and one of the epicyclic frequencies $m:n$, where $m, n =1,2,3$ don't exist when we have the ordering $\omega_0<\omega_r$ or $\omega_0<\omega_\theta$.

The location of the described resonances as a function of the spin parameter is illustrated in figs. $\ref{fig:res1}$-$\ref{fig:res2}$. For the co-rotating orbits the resonances are excited in the close vicinity of the wormhole throat, i.e. in the region with  a very strong gravitational field. Moreover, this behavior is  observed not only for rapidly rotating wormholes, but for a wide range of values of the wormhole  spin. Thus, the quasiperiodic oscillations in wormhole spacetimes can be a valuable probe for the strong gravity regime. Another characteristic feature is that for a fixed value of the spin parameter the same type of resonance occurs for several different radii. In the case of the Kerr black hole the resonant curves are monotonic and such behavior is excluded. This phenomenon is particularly interesting since the radius where the resonance is excited is connected with the properties of the physical process causing it. Thus, for wormhole spacetimes we can have the same type of resonance excited simultaneously at different regions in the accretion disk probably caused by different physical processes.

\section{Conclusion}

Wormholes are one of the major predictions of the gravitational theories, which is still awaiting experimental confirmation. Therefore, it is important to be familiar with their characteristics in different observable phenomena. In this work we study how we can interpret the high-frequency quasi-periodic oscillations from the accretion disk within the resonance models if we assume that the central compact object represents a wormhole instead of the Kerr black hole. For the purpose we consider the traversable wormhole geometry derived by Teo, which describes any stationary and axisymmetric completely regular wormhole solution within classical or semi-classical gravity. We perform a systematic study of the existence and  stability of the timelike circular geodesics in the equatorial plane. As a result we  derive analytical expressions for the epicyclic frequencies, which govern the evolution of small deviations from the circular motion, which are valid for a general class of traversable wormholes with integrable geodesic equations. We see that for large classes of wormholes the vertical epicyclic frequency is always positive, ensuring that circular orbits are always stable with respect to small perturbations in vertical direction. In this respect wormholes are similar to the Kerr black hole since the stability is determined only by  the radial epicyclic frequency.

In other aspects the quasi-circular equatorial motion in wormhole spacetimes shows significant differences. A major distinction is that  the epicyclic and orbital frequencies can obey different types of ordering in the various regions of the parametric space. In contrast, for the Kerr black hole they maintain a constant relation for any radius and spin parameter. This property enables the manifestation of a richer class of resonant phenomena in wormhole spacetimes opening new possibilities for the explanation of the observed quasi-periodic oscillations from the accretion disk. In particular, lower order parametric and forced resonances are possible, which will lead to stronger observable signals. For a wide range of spin parameters resonances can be excited in the close neighbourhood of the wormhole throat, probing the region of strong gravitational interaction. In addition, the same type of resonance can take place simultaneously at several different radial distances, which can put some restrictions on the physical processes in the accretion disk giving origin to the resonant phenomena.

\section*{Acknowledgments}
We gratefully acknowledge support by the DFG Research Training Group 1620 ``Models of Gravity''  and the COST Actions CA16214 and CA16104.  P.N. is partially supported by the Bulgarian NSF Grant  KP-06-H38/2.

\end{document}